\pdfoutput=1

\documentclass[authorversion,nonacm,sigplan,screen]{acmart}
\usepackage{pifont}

\newcommand{\xmark}{\ding{55}}

\usepackage[ruled, linesnumbered]{algorithm2e}
\usepackage{subcaption}
\usepackage{multirow}

\usepackage{flafter}

\usepackage{tikz}
\usepackage{xcolor}
\newcommand*\circled[1]{\tikz[baseline=(char.base)]{
  \node[shape=circle,fill,inner sep=2pt] (char)
{\textcolor{white}{\textbf{#1}}};}}

\copyrightyear{2023}
\acmYear{2023}
\setcopyright{none}
\acmConference[PPoPP '23]{The 28th ACM SIGPLAN Annual Symposium on Principles and Practice of Parallel Programming}{February 25-March 1, 2023}{Montreal, QC, Canada}
\acmBooktitle{The 28th ACM SIGPLAN Annual Symposium on Principles and Practice of Parallel Programming (PPoPP '23), February 25-March 1, 2023, Montreal, QC, Canada}
\acmDOI{10.1145/3572848.3577480}
\acmISBN{979-8-4007-0015-6/23/02}

\newcommand{\bdm}{BioDynaMo}

\newcommand{\result}[1]{#1}
\definecolor{dora}{rgb}{0.80, 0.60, 0.00}

\begin{document}

\title[High-Performance and Scalable Agent-Based Simulation with \bdm{}]{High-Performance and Scalable \\Agent-Based Simulation with \bdm{}}

\author{Lukas Breitwieser}
\orcid{0000-0003-2265-8615}
\affiliation[obeypunctuation=true]{\institution{CERN},
\country{Switzerland}
}
\affiliation[obeypunctuation=true]{\institution{ETH Zurich},
\country{Switzerland}
}
\authornote{lukas.breitwieser@gmail.com}

\author{Ahmad Hesam}
\orcid{0000-0001-7331-1000}
\affiliation[obeypunctuation=true]{\institution{Delft University of Technology},\\
\country{The Netherlands}
}

\author{Fons Rademakers}
\orcid{0000-0002-3571-9635}
\affiliation[obeypunctuation=true]{\institution{CERN},
\country{Switzerland}
}

\author{Juan G\'{o}mez Luna}
\orcid{0000-0002-6514-1571}
\affiliation[obeypunctuation=true]{\institution{ETH Zurich},
\country{Switzerland}
}

\author{Onur Mutlu}
\orcid{0000-0002-0075-2312}
\affiliation[obeypunctuation=true]{\institution{ETH Zurich},
\country{Switzerland}
}
\authornote{omutlu@gmail.com}

\renewcommand{\shortauthors}{Breitwieser, et al.}

\begin{abstract}
Agent-based modeling plays an essential role in gaining insights into biology,
  sociology, economics, and other fields.
However, many existing agent-based simulation platforms are not suitable for
  large-scale studies due to the low performance of the underlying simulation
  engines.
To overcome this limitation, we present a novel high-performance simulation
  engine.

We identify three key challenges for which we present the following solutions.
First, to maximize parallelization, we present an optimized grid to search for neighbors and parallelize the merging of thread-local results.
Second, we reduce the memory access latency with a NUMA-aware agent iterator, agent sorting with a space-filling curve, and a custom heap memory allocator.
Third, we present a mechanism to omit the collision force calculation under certain conditions.

Our evaluation shows an \result{order of magnitude} improvement over Biocellion, \result{three orders of magnitude} 
speedup over Cortex3D and NetLogo, and the ability to simulate \result{1.72 billion} agents on a single server. 

Supplementary Materials, including instructions to reproduce the results, are available at:
\url{https://doi.org/10.5281/zenodo.6463816}
 \end{abstract}



\maketitle

\pagebreak\section{Introduction}

Agent-based modeling (ABM) allows to simulate complex dynamics in a wide range
  of research fields.
ABM has been used to answer research questions in biology
  \cite{metzcar_review_2019, zubler_simulating_2013, hunter_taxonomy_2017},
  sociology \cite{epstein_growing_1996}, economics
  \cite{tesfatsion_chapter_2006}, technology \cite{niazi_agent-based_2009},
  business \cite{rand_agent-based_2011}, and more fields
  \cite{macal_introductory_2014}.
\emph{Agents} are individual entities that, among others, can represent
subcellular
structures to simulate the growth of a neuron, a cell to investigate cancer
development, or a person to simulate the spread of infectious diseases
\cite{breitwieser-bdm}.
The actions of an agent are defined through instances of class \emph{behavior}.
To stay with the examples from before, possible behaviors are neurite
  bifurcation, uncontrolled cell division, or infection.

Agent-based models are developed in an iterative way, during which an initial model 
  is increasingly refined until it matches with observed data 
  \cite{thorne_combining_2007, roberts_mathematical_2016}. 
Model parameters that cannot be derived from the literature are determined
  through optimization. 
An optimization algorithm generates a parameter set, executes the model, and 
  evaluates the error with respect to observed data until the error converges to
  a local or global minimum.
This loop might also contain an uncertainty analysis to evaluate the robustness of 
  a solution \cite{marino_methodology_2008}.
Consequently, the model must be simulated many times. 

The simulation engine's performance limits the scale of the model and determines how often the 
  model can be simulated.
Thus, performance is a key issue for simulating models on extreme scales that might one 
  day be able to simulate all 86 billion neurons in the brain \cite{azevedo_equal_2009}. 
It is also crucial for smaller-scale simulations to explore vast parameter space, 
  analyze parameter uncertainty, repeat the simulation often enough to reach statistical 
  significance, and develop models rapidly.

To achieve these goals, we present a novel simulation engine called \bdm{},
which is optimized for high performance and scalability.
During its development, we identify the following three main performance
  challenges for agent-based simulations.

\textbf{Challenge 1:}
To fully utilize systems with high processor core counts, the parallel part of
  the simulation engine has to be maximized (see Amdahl's law
  \cite{amdahl_validity_1967}).
Although it is easy to parallelize the loop over all agents
  (Algorithm~\ref{algo:main}), our benchmarks revealed two operations whose level
  of parallelization has a significant performance impact.
First, building the environment index, which is used to determine the neighbors
  of an agent.
The literature describes various radial-neighbor search algorithms with
  different design trade offs between build and search performance.
Second, combining thread-local results at the end of each iteration.
In general, attention must also be paid to seemingly minor things, such as
  resizing a large vector, which by default is initialized by a single thread.

\textbf{Challenge 2:}
ABMs are predominantly memory-bound due to two reasons.
First, the behavior of agents often has low arithmetic intensity.
Second, ABM can be very dynamic.
During a simulation, agents move through space, change their behavior, and are
  created and destroyed.
Consequently, the neighborhood of an agent changes continuously, leading to an
  irregular memory access pattern and poor cache utilization.
This results in large data movement between the main memory and the processor
  cores.

\textbf{Challenge 3:}
Under certain conditions, the expensive calculation of mechanical forces
  between agents is redundant (Section~\ref{sec:static-agents}).
These forces are for example used in tissue models to determine the
  displacement of agents.
The challenge is identifying those agents for which the pairwise force
  calculation can be safely omitted.

\bdm{} addresses these challenges with the following new optimizations.
To maximize the parallelization (Challenge~1), we develop an optimized uniform
  grid to search for agent neighbors and fully parallelize the addition and
  removal of agents.
We address the data movement bottleneck (Challenge~2) in software by (i)
  optimizing the iteration over all agents on systems with non-uniform memory
  architecture, (ii) sorting agents and their neighbors to improve the cache hit
  rate and minimize access to remote DRAM, and (iii) introducing a pool memory
  allocator.
To avoid redundant mechanical force calculations (Challenge~3), we add a
  mechanism to detect agents for which we can guarantee that the resulting force
  will not move the agent.

These mechanisms make \bdm{} \result{nearly an order of magnitude} more efficient than Biocellion 
  and three orders of magnitude faster than Cortex3D and NetLogo.
The performance improvements account for a median speedup of
  \result{159$\times$} compared to \bdm{}'s standard implementation with all
  optimizations turned off.
As a result, \bdm{} is able to simulate \result{1.72 billion} agents on one server. 
The main contributions of this paper are as follows.

\begin{itemize}
  \item We present a novel high-performance agent-based simulation engine.
The engine can be used in many domains due to its modular software design and
          features a specialization for neuroscience, capable of simulating the
          development of neurons.

\item We present six optimizations to maximize performance (Section~\ref{sec:opt:maximize-parallelization}--\ref{sec:static-agents}).
These insights are transferable and can be used to improve the performance of
          other agent-based simulators.

  \item We present an in-depth evaluation of \bdm{}'s performance using five
        different simulations (Section~\ref{sec:evaluation}).
This comprehensive analysis provides insights for users of \bdm{} into which
          parameters yield the best performance and hints for developers of future
          agent-based simulation tools.

\end{itemize}

\section{\bdm{}'s Simulation Engine}
\label{sec:design-overview}

This Section gives an overview of \bdm{} and its components.
\bdm{} is written in C++, uses OpenMP \cite{openmp} for shared-memory
  parallelism, and is available under the Apache 2.0 open-source license.

Breitwieser et al. \cite{breitwieser-bdm} describe the user-facing features of the \bdm{} platform and detail its modular software design and ease-of-use by means of three use cases in the domains of neuroscience, epidemiology, and oncology. 

\bdm{} is a hybrid framework able to utilize multi-core CPUs and GPUs. 
This paper focuses on the CPU version, which has two major advantages.
First, the CPU version can simulate many more agents than a GPU version. 
The reason is that GPUs have typically significantly smaller memory than CPUs. 
For example, our benchmark hardware has 12$\times$ more memory than the current flagship GPU from NVidia, the A100 \cite{nvidia2020a100}. 
Second, the CPU version improves the usability and flexibility for our broad user community, 
who often only have a Matlab \cite{matlab} or R \cite{r} coding background.
In \bdm{}, users create simulations by writing C++ code.
A GPU-only version would require users to write CUDA code to define new agents, behaviors, and other user-defined components.
Therefore, \bdm{} only offloads computations to the GPU, transparently to the user \cite{hesam_gpu}.

The main objects in agent-based simulations are agents, behaviors, and
  operations.
Agents (e.g., a cancer cell) have attributes that are updated through behaviors
  and operations.
Behaviors (e.g., uncontrolled cell division) are functions that can be assigned
  and removed from an agent and give users fine-grained control over the actions
  of an agent.
In contrast, \emph{Agent operations} are executed for each agent.
For example, to calculate the mechanical forces between agents and execute all
  individual behaviors of an agent.
The second type of operation, called \emph{standalone operation} is executed
  once per iteration to perform a specific task (e.g., visualization).
A characteristic property of agent-based simulation is local interaction.
\bdm{} provides a common interface for different neighbor search algorithms
  called \emph{environment}.
Besides the uniform grid detailed in Section~\ref{sec:grid}, \bdm{} features a
  kd-tree based on nanoflann \cite{nanoflann} and octree based on the publication
  of Behley et al.
\cite{unibn}.

The agent-based simulation algorithm (Algorithm~\ref{algo:main}) comprises two
  steps.
First, users have to define the starting condition of the model (L1) in which
  agents, behaviors, operations, and any other resource are created.
Second, the simulation engine executes this model for a number of iterations
  (L2--19).
The engine executes all agent operations for each agent (L7--12) and all
  standalone operations (L12--14).
Standalone operations can be further separated into operations that must be
  executed at the beginning of the iteration (e.g., to update the environment
  index [L3-5]) or the end (e.g., visualization [L16-18]).
There are two barriers synchronizing threads (L6 and L15).

\makeatletter
\newcommand{\removelatexerror}{\let\@latex@error\@gobble}
\makeatother

\newcommand{\myalgorithm}{\begingroup
\removelatexerror \begin{algorithm}[H]
\caption{Simulation algorithm}
  \label{algo:main}
\scriptsize

  \SetKwFor{ParallelFor}{parallel for}{do}{end}
  \SetKwFunction{Wait}{wait}
  \SetKwFunction{ModelInitialization}{ModelInitialization}
  \SetKwFunction{PostProcessing}{PostProcessing}
  
  \ModelInitialization{}

  \For{$i\in iterations$}{
    \For{$op\in pre\_standalone\_operations$}{
      $op()$\;
    }

    \Wait{}

    \ParallelFor{$a\in agents$}{
      \For{$op\in agent\_operations$}{
        $op(a)$\;
      }
    }
    \For{$op\in standalone\_operations$}{
      $op()$\;
    }

    \Wait{}

    \For{$op\in post\_standalone\_operations$}{
      $op()$\;
    }
  }

\end{algorithm}
\endgroup}

\begin{figure}[!h]
  \centering
  \begin{minipage}{.75\linewidth}
        \myalgorithm
  \end{minipage}
\end{figure}

\section{Maximize Parallelization}
\label{sec:opt:maximize-parallelization}

\subsection{Grid-based Neighbor Search}
\label{sec:grid}

Determining the neighbors of an agent is a pre-condition for all agent
  interactions.
For example, the infection behavior in an epidemiological model requires
  information if any of the immediate neighbors is infected.
In this context, it is essential to find neighbors fast and efficiently and
  minimize the build time of the required index.
Building an index in every iteration has a high cost, as shown in the
  evaluation section.
We exploit the fact that the interaction radius is known at the beginning of
  the iteration.
For this fixed-radius search problem, a grid-based solution is a good choice
  because the box of an agent can be determined in constant time using the
  agent's position \cite{unibn}.
This is confirmed by our evaluation in Section~\ref{sec:eval:environment}.
The build stage in which all agents are assigned to a box can be easily
  parallelized.
In the search stage, the grid determines all neighbors by iterating over all
  agents in the same box and the surrounding boxes.
In 3D space, we consider the 3x3x3 cube of boxes surrounding and including the query box. 
All agents inside a box are stored in an array-based linked list.
The box only needs to store the start index and the number of elements it
  contains.
To avoid zeroing all boxes at the beginning of the build stage, we add a
  timestamp attribute to each box, updated whenever an agent is added.
Consequently, we can determine that a box is empty if the simulation and box
  timestamp is different.
Therefore, we can build the grid in $O(\#agents)$ time instead of $O(\#agents +
    \#boxes)$, which is relevant for large simulation spaces that are not fully
    populated.

The array-based linked list uses the same agent indices as in the
  \texttt{ResourceManager}.
The \texttt{ResourceManager}, an essential class in the simulation engine,
  stores raw agent pointers and offers functions to add, remove, get, and iterate
  over agents.
Thus, it also benefits from the memory layout optimization presented in
  Section~\ref{sec:load-balancing}.
This optimization reduces the distance in memory of agents that are close in space. 
Consequently, linked list elements will be closer to each other, improving the 
  cache hit rate of traversing the linked list during the search stage of the grid.
The described grid implementation can be found in the class \texttt{UniformGridEnvironment}.
 
\subsection{Adding and Removing Agents in Parallel}
\label{sec:parallel-remove}

To maximize the theoretically achievable speedup described in Amdahl's law
  \cite{amdahl_validity_1967}, we maximize the parallel part of the simulation by
  parallelizing the addition and removal of agents.
By default, \bdm{} stores a thread-local
  copy of additions and removals and commits them to the \texttt{Resource\-Manager}
  at the end of each iteration.

\begin{figure}[tb]
  \centering
\includegraphics[width=0.9\linewidth]{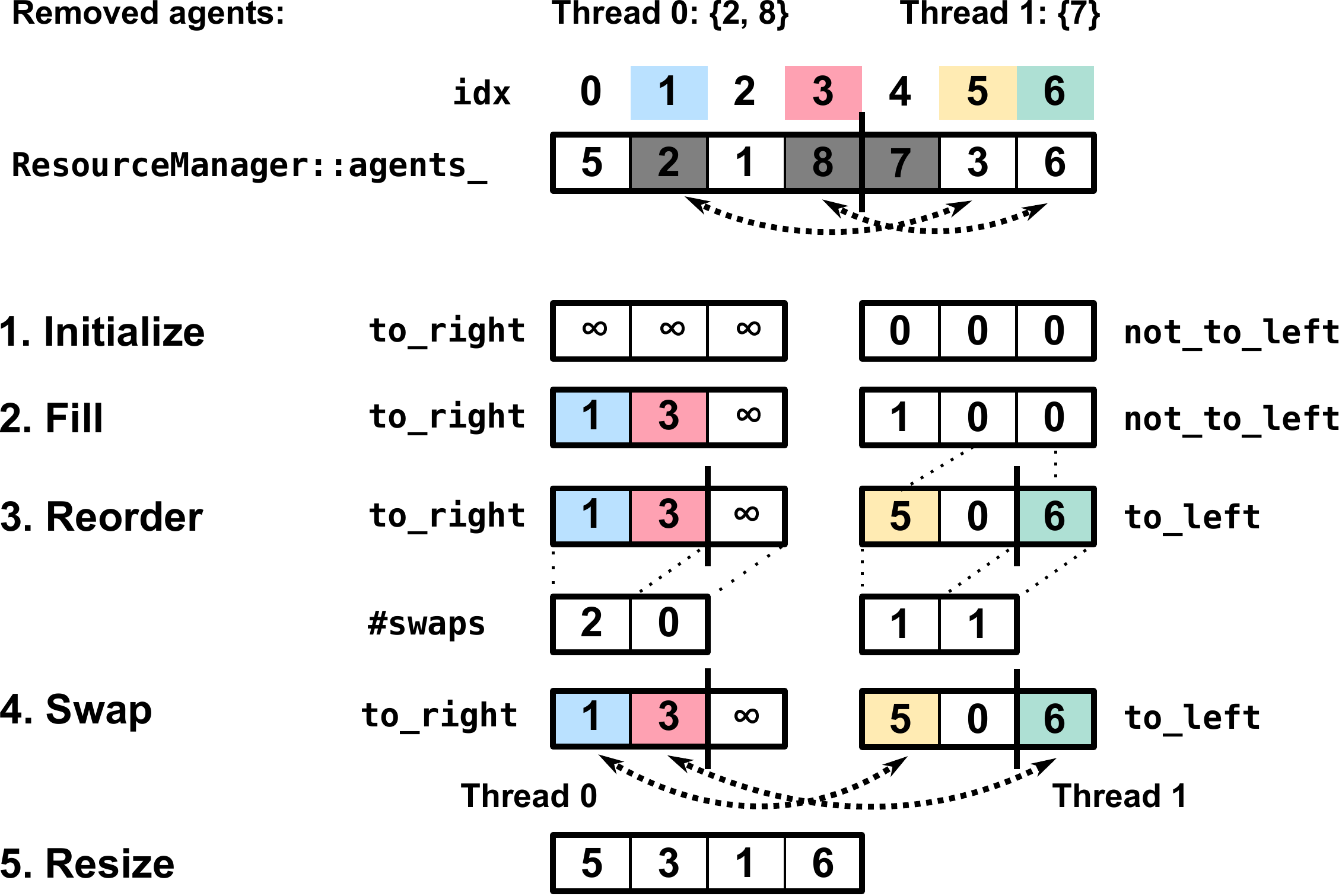}
\caption{Parallel agent removal mechanism}
  \label{fig:parallel-remove}
\end{figure}

Additions are trivial; the engine determines the total number of additions,
  grows the data structures in the \texttt{Resource\-Manager}, and adds the agent
  pointers in parallel.
In contrast, the parallelization of removals is a more elaborate process
  because we disallow empty vector elements in the \texttt{ResourceManager}.
If the simulation engine has to remove an agent stored in the middle of the
  vector, it must swap it with the last element before shrinking it.
The following algorithm aims at performing the necessary swaps and updates in
  dependent data structures in parallel.
Figure~\ref{fig:parallel-remove} illustrates the parallelized algorithm
  simplified for a single NUMA domain.
 This example assumes a simulation with seven agents represented with identifier
  1--7 and two threads, which remove three agents from the simulation.
These agents are highlighted with a grey background.
The other colors serve as a visual aid to track the agents that must be
  swapped.

The algorithm comprises five main steps.
First, the algorithm determines the total number of removed agents, calculates
  the new size of the vector ($old\_size - removed\_agents$), and initializes two
  auxiliary arrays.
The size of the auxiliary arrays equals the number of removed agents.
The new vector size is indicated by the vertical line between indexes three and
  four in Figure~\ref{fig:parallel-remove}.

Second, each thread iterates over its vector of removed agents and fills the
  auxiliary arrays.
If the agent is stored to the left of the new size index, it must be moved to
  the right.
Therefore the algorithm inserts the element index into the array $to\_right$.
If the agent is stored to the right of the new size, we insert a one into array
  $not\_to\_left$ at the index: $idx - new\_size$.
The maximum index used to access elements in the auxiliary array is smaller than 
  the number of removed agents and is independent of the number of remaining agents.

Third, we partition the two auxiliary arrays into blocks corresponding to the
  total number of threads.
A thread iterates over its auxiliary block, moves entries to the beginning if
  they indicate a swap, and stores a counter in the \texttt{\#swaps} array.
For the $to\_right$ array, the algorithm skips or overwrites elements with the
  value $UINT\_MAX$.
In this step, the semantic of the $not\_to\_left$ array changes to $to\_left$.
Thus, the algorithm looks for all zeros in the array, replaces them with the
  value $array\_index + new\_size$, and moves them to the beginning of the block.

In the fourth step, we can finally perform the swaps.
The algorithm calculates the prefix sum of the two $\#swaps$ arrays and
  partitions the swaps among all threads.
Each thread can determine the indices based on the prefix sum of $\#swaps$.

In the last step, the algorithm completes the removal by shrinking the vector
  to $new\_size$.

This algorithm requires $O(removed\_agents)$ time and space and parallelizes
  steps 1--4.
 
\section{Optimize Memory Layout and Data Access Pattern}
\label{sec:opt:memory-layout}

\subsection{NUMA-Aware Iteration}
\label{sec:numa-iteration}

\bdm{} supports systems with multiple NUMA domains.
We add a mechanism to match threads with agents from the same NUMA domain to
  minimize the traffic to remote DRAM because OpenMP does not provide this
  functionality.

\begin{figure}[tb]
  \centering
\includegraphics[width=\linewidth]{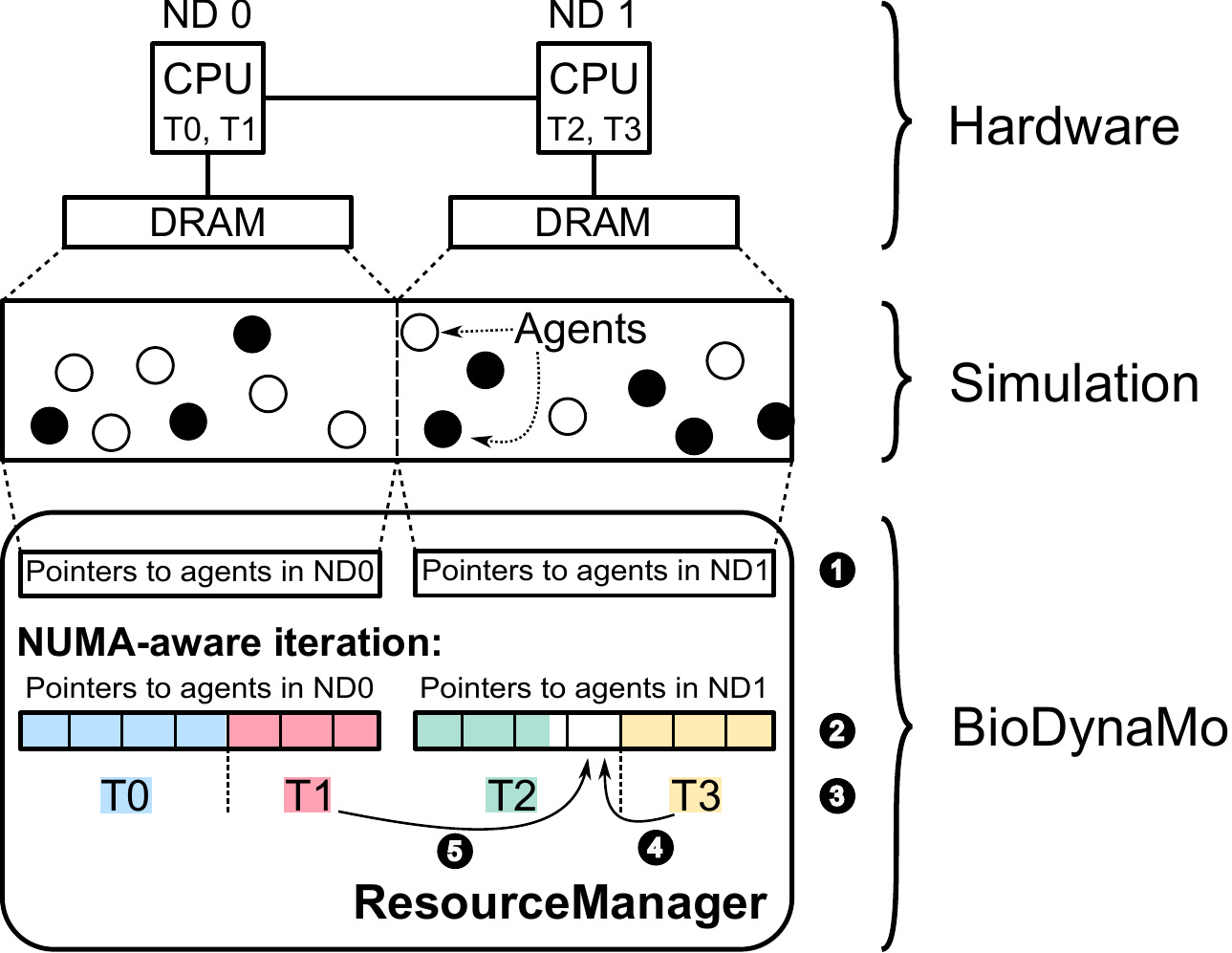}

\caption{NUMA-aware iteration}
\label{fig:numa-iteration}
\end{figure}

Figure~\ref{fig:numa-iteration} shows a server with two NUMA 
  domains (\texttt{ND0}, \texttt{ND1}) corresponding to two CPUs with two threads each (T0 \& T1, T2 \& T3).
 The CPUs have a local DRAM with shorter memory access latency than the remote DRAM.
The agents in the simulation are balanced between the two NUMA domains (see Section~\ref{sec:load-balancing}).
The \texttt{ResourceManager} maintains a vector of agent pointers for each NUMA
  domain \circled{1}.
To iterate over agents in a NUMA-aware way, \bdm{} first partitions these vectors
  into blocks of agent pointers of the same size \circled{2}.
Second, these blocks are partitioned among the threads from the matching NUMA domain \circled{3}.
Threads process the assigned blocks in parallel.
Figure~\ref{fig:numa-iteration} shows processed blocks with a background color of the corresponding thread.

We implement a two-level work-stealing mechanism
  to avoid imbalanced execution times across threads.
First, a thread can steal a block from a different thread from the \emph{same} NUMA
  domain (e.g., \circled{4}).
Second, if the thread's NUMA domain has already finished all work, the thread
  can steal work from a different NUMA domain (e.g., \circled{5}).

\subsection{Agent Sorting and Balancing}
\label{sec:load-balancing}

To accelerate the memory-bound simulations, we must increase the cache hit
  ratio and load balance the agents among NUMA domains to minimize remote DRAM
  accesses.
In Section~\ref{sec:numa-iteration} and Figure~\ref{fig:numa-iteration}, we assume this is already the case.
This section presents an efficient algorithm to achieve this goal by sorting the agents' memory
  locations and preserving the neighborhood relations in 3D.
Preserving the neighborhood relation and reducing the dimensionality is the
  main characteristic of space-filling curves (e.g., Morton order
  \cite{morton1966computer} or Hilbert curve \cite{hilbert_ueber_1891}).

We compared the performance of the Morton order with the Hilbert curve using an oncological 
  simulation \cite{breitwieser-bdm} and observed a negligible performance improvement of 0.54\% from using the Hilbert curve. 
Higher costs to decode the Hilbert curve offset small gains for the agent operations. 
Therefore, we use the Morton order because it results in simpler code.

\begin{figure}[tb]
  \centering
\includegraphics[width=\linewidth]{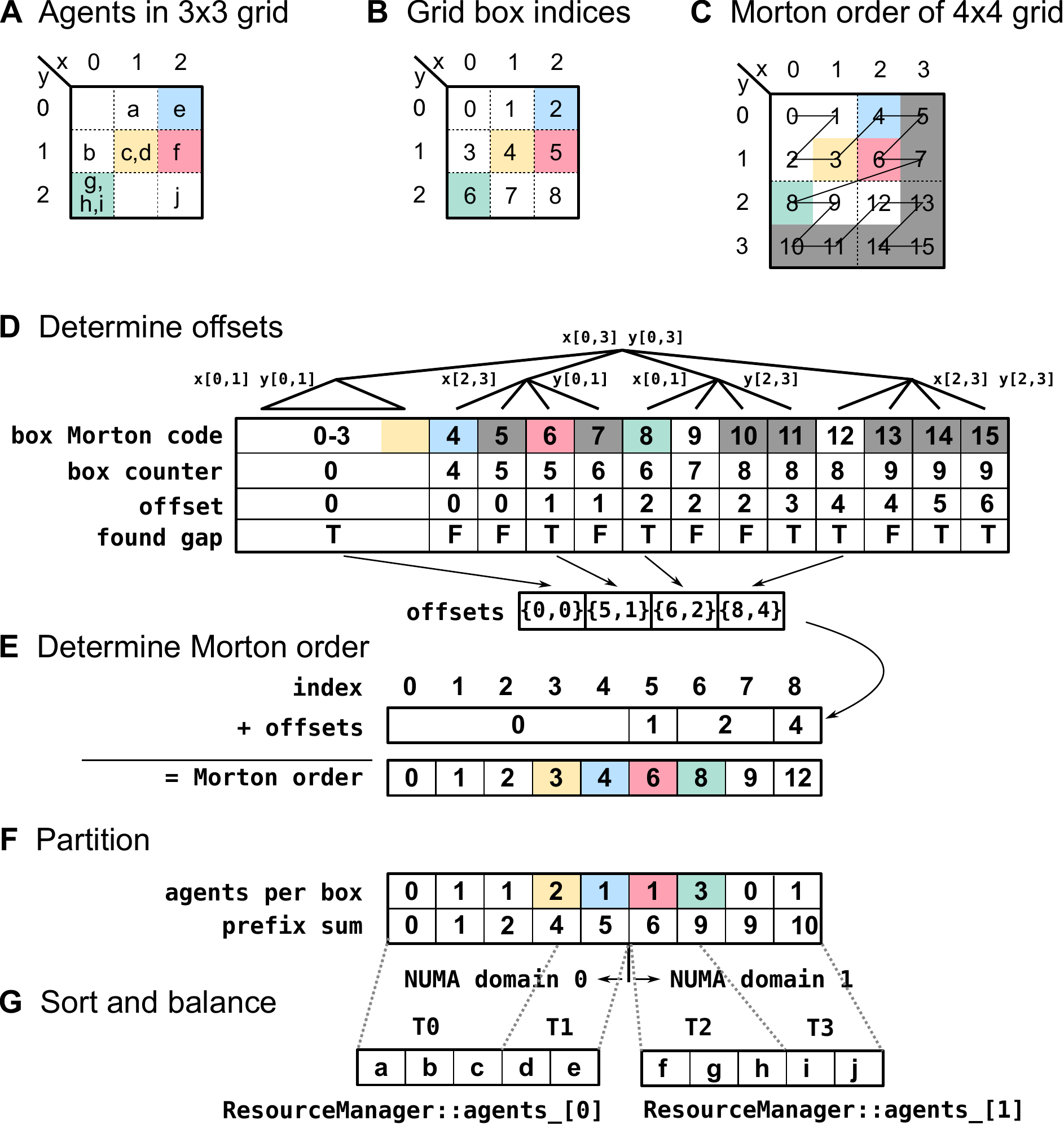}
\caption{Agent sorting and balancing mechanism}
\label{fig:load-balancing-design}
\end{figure}

Figure~\ref{fig:load-balancing-design} and the following description present
  the algorithm in 2D space, but the same principles apply in 3D.
In \bdm{}, the neighborhood information is stored in the implementation of the
  environment interface.
 Since the uniform grid environment performs best, as shown in the evaluation
  section (Section~\ref{sec:eval:environment}), we utilize its characteristics to
  achieve fast sorting and balancing.
We assume the following scenario.
Agents are stored in a 3$\times$3 uniform grid (A).
The simulation runs on a system with two NUMA domains and two threads per
  domain, resulting in four threads.
The grid boxes are stored in a flattened array.
Figure~\ref{fig:load-balancing-design}B shows the box indices for x and y
  coordinates.
We apply a Morton order space-filling curve
  \cite{morton1966computer} on the uniform grid (C).
Our goal is to sort the agents in the nine boxes in increasing Morton order.
The Morton order is only contiguous for quadratic simulation spaces where the
  length is a power of two.
Therefore, C shows a 4$\times$4 grid.
For the 3$\times$3 simulation space, there are gaps between Morton code four
  and six, six and eight, and nine and twelve.

The algorithm comprises three main steps.
First, the algorithm determines the sequence of boxes in Morton order (D, E).
Second, the algorithm partitions the boxes into segments that balance agents
  among NUMA domains and threads (F).
Third, the algorithm stores the agents in their new position in the resource
  manager (G).
In Figure~\ref{fig:load-balancing-design}, we selected four boxes and colored
  them in red, green, blue, and yellow to quickly find the corresponding entries
  for contained agents, box index, and Morton code.
The boxes outside the simulation space have a grey background.

In the first main step (D, E), we determine all gaps between grid boxes in
  simulation space (D) to avoid a costly sorting operation or iteration over all
  $N\times N$ boxes, where $N$ is the next higher power of two of $max(x, y)$.
We exploit the fact that the Morton order corresponds to a depth-first
  traversal of a quadtree, in which each box is a leaf in the tree.
Leaves whose boxes are outside the simulation space are considered empty.
Similarly, an inner node is empty if all of its corresponding leaves are
  outside the simulation space.
If all the corresponding boxes of an inner node are inside the simulation space
  (i.e., the inner node has a perfect subtree), we say the inner node is
  complete.
The quadtree is only an abstraction and does not need to be constructed.
It is only necessary to store the current traversal path, which requires
  $O(log(\#boxes))$ space.

The mechanism in D uses three auxiliary variables: box counter,  offsets, and
  \texttt{found\_gap}, which are initialized to zero, zero, and true.
The matrix in Figure~\ref{fig:load-balancing-design}D shows the three variables
  before the update of the current traversal step.
The algorithm traverses the tree depth-first and continues to the next deeper
  tree level only if the current node is neither complete nor empty.
In this case, the variables are not changed.
If a complete inner node or leave inside the simulation space is found and the
  \texttt{found\_gap} variable is true, the algorithm adds an entry with the
  current box counter and offset values in the offsets array and clears
  \texttt{found\_gap}.
Afterward, the box counter variable is incremented by the number of leaves in
  its subtree or one if it was a leave, irrespectively of the former value of
  \texttt{found\_gap}.
Empty nodes or leaves are handled similarly.
The offset variable is incremented by the number of empty leaves in the subtree
  of an empty node or one if it is an empty leave.
Additionally, the \texttt{found\_gap} variable is set to true.
The algorithm keeps track of the x and y intervals to calculate (in constant
  time) the number of leaves in a subtree and determine if an inner node is
  entirely, partially, or not inside the simulation space.

With the already sorted offsets array, the Morton order can be determined in
  linear time by iterating over all indices and adding the corresponding offset
  (E).

In the second main step (F), the algorithm iterates over all boxes in Morton
  order and fills an auxiliary array with the number of agents in each box.
Afterward, the algorithm calculates the prefix sum of the auxiliary array in a
  parallel work-efficient manner \cite{ladner_parallel_1980} and partitions the
  total number of agents in the simulation such that each NUMA domain receives a
  share corresponding to its number of threads.
Inside a NUMA domain, the agents are further partitioned such that each thread
  in this domain receives an equal share.

In the third main step (G), the threads copy the agents and store the pointer
  in the new position in the resource manager.
The simulation engine can immediately free obsolete agents' memory or delete
  all old copies after the step is finished.
The latter requires more memory but might improve performance due to a more
  optimal memory layout in conjunction with the \bdm{} memory allocator
  (Section~\ref{sec:mem-allocator}).

The presented algorithm runs in $O(\#agents + \#boxes)$ time and space and
  parallelizes steps E--G.
 
\subsection{\bdm{}
Memory Allocator} \label{sec:mem-allocator} 

  To improve the performance of the simulation engine, we present a custom
  dynamic memory allocator which improves the memory layout of the most
  frequently allocated objects in a simulation: agents and behaviors.
Our solution builds upon pool allocators due to their constant time allocation
  performance.
Pool allocators divide a memory block into equal-sized elements and store
  pointers to free elements in a linked list.

We create multiple instances of these allocators because they can only return
  memory elements of one size.
As a result, agents and behaviors with distinct sizes are separated and stored
  in a columnar way.
We separate the pool allocator into multiple NUMA domains (class
  \texttt{NumaPoolAllocator}) to fully control where memory is allocated.
The \texttt{Numa\-Pool\-Allocator} has a central free-list and thread-private ones
  to minimize synchronization between threads.
List nodes, which correspond to free memory locations, can be migrated between
  thread private and the central list, which is essential to avoid memory leaks.
Migrations are triggered if a thread-private list exceeds a specific memory
  threshold.
Lists minimize these migrations, and thus thread synchronization, by
  maintaining additional skip lists.
These skip lists support additions and removals of a large number of elements
  in constant time.

Memory is allocated in large blocks with exponentially increasing sizes
  controlled by the parameter \texttt{mem\_\-mgr\_\-growth\_\-rate}.
The initialization of these memory blocks, which includes list node generation,
  is performed on-demand in smaller segments to minimize the required worst-case
  allocation time.

\begin{figure}[tb]
  \centering
\includegraphics[width=\linewidth]{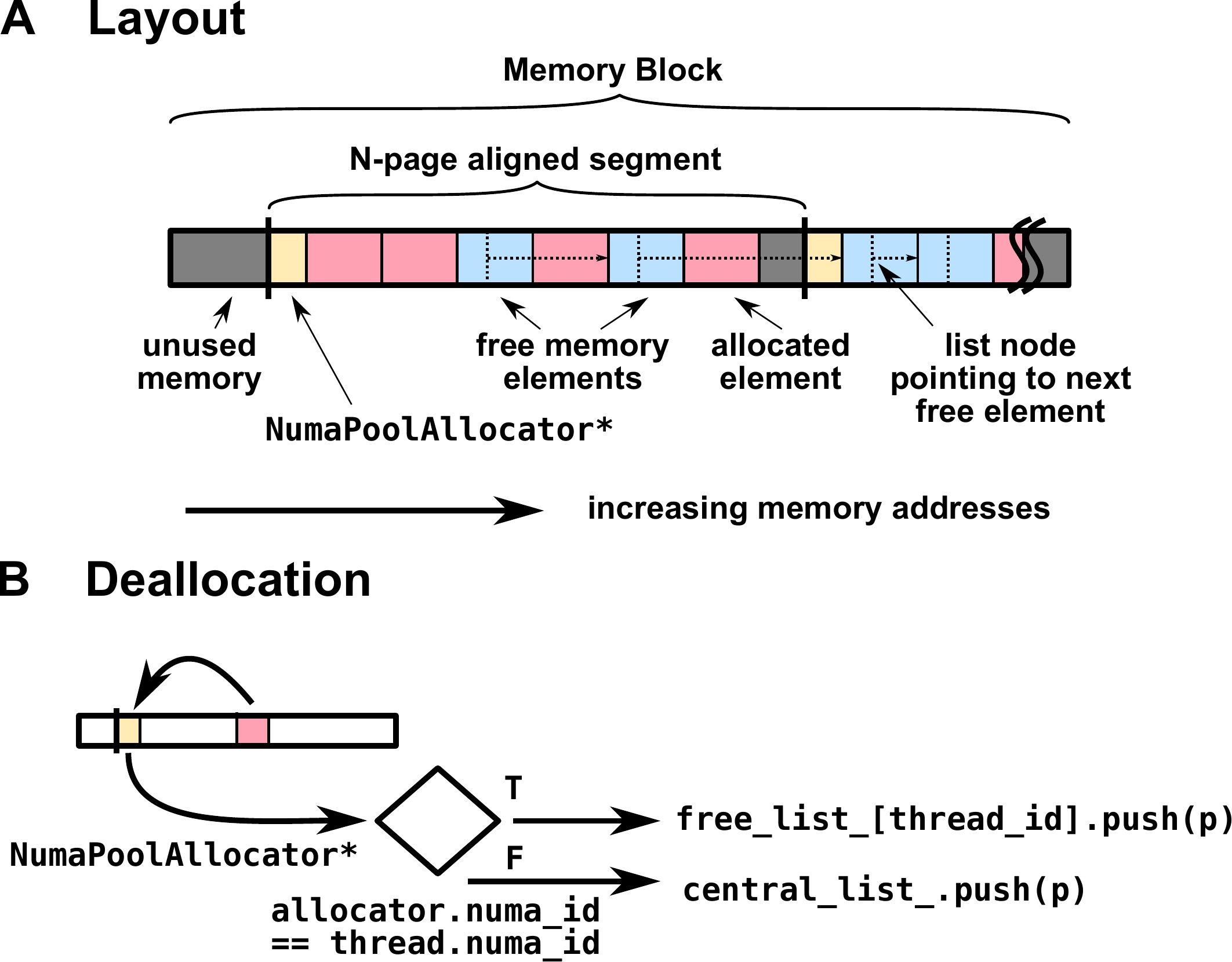}

\caption{\bdm{}'s memory allocator}
\label{fig:mem-mgr}
\end{figure}

Every allocated memory block is divided into N-page aligned segments
  (Figure~\ref{fig:mem-mgr}A), where N can be set with parameter
  \texttt{mem\_mgr\_aligned\_pages\_shift}.
 The linked list nodes are stored inside free memory elements and do not require
  extra space.
At the beginning of each N-aligned segment, we write the pointer to the
  corresponding \texttt{NumaPoolAllocator} instance.
Therefore, allocated memory elements can obtain this pointer in constant time,
  based on their memory address. This solution enables constant time deallocations (see Figure~\ref{fig:mem-mgr}b) but wastes memory in three
  ways.
First, memory blocks are allocated using \texttt{numa\_alloc\_onnode} (libnuma
  \cite{libnuma}).
This function does not return N-page aligned pointers and causes unusable
  regions at the beginning and the end, which sum up to $N * page\_size$ bytes.
Second, there might not be enough space to place a whole element at the end of
  an N-page aligned segment.
Elements must not cross N-page aligned borders because it would overwrite the
  necessary metadata.
Third, the metadata requires the size of a pointer, which is eight bytes on
  64-bit hardware.
The amount of wasted memory is bounded by the following equation: $N *
    page\_size + element\_size + metadata\_size$.
Despite this memory overhead, our performance evaluation shows that the \bdm{}
  pool allocator uses \result{on average less memory} than ptmalloc2 and jemalloc
  \cite{evans2011scalable}.
Another side effect of this design choice is that the allocation size is
  limited by $N * page\_size - metadata\_size$.

\section{Omit Collision Force Calculation}
\label{sec:static-agents}

The most time-consuming operation in the tissue models presented in
  Section~\ref{sec:benchmarks} is the calculation of the displacement of agents
  based on all mechanical forces.
For this purpose, the simulation engine has to calculate pairwise collision
  forces between agents and their neighbors implemented in the class
  \texttt{InteractionForce}.
By default, \bdm{} uses the force calculation method detailed in the Cortex3D paper \cite{zublerdouglas2009framework}.
We observe that simulations can contain a significant amount of regions where
  agents do not move.
Neural development simulations (see Section~\ref{sec:benchmarks}), for example,
  might only have an active growth front, while the remaining part of the neuron
  is unchanged.

Therefore, we present a mechanism to detect agents for which it is safe to skip
  the expensive force calculation.
We call these agents static.

The following four conditions must be fulfilled in the last iteration: (i) the
  agent and none of its neighbors moved (ii) the agent's and neighbors'
  attributes did not change in a way that could increase the pairwise force
  (e.g., larger diameter), or the resulting displacement, (iii) new agents were
  not added within the interaction radius of the agent, and (iv) there is maximum
  one neighbor force which is non-zero.

The detection mechanism is closely tied to the \texttt{Interaction\-Force}
  implementation (see \cite{breitwieser-bdm}), as condition two implies, and
  might have to be adjusted if a different force implementation is used.

Condition four is needed because we want to allow agents to shrink and to be
  removed from the simulation without setting the agents in this region to not
  static.
Consequently, we have to ensure that two or more neighbor forces did not cancel
  each other out in the previous iteration.

The simulation engine monitors if any of the conditions are violated for each
  agent and sets the affected agents to not static.
In this process, a distinction has to be made whether the changed attribute
  affects only the current agent or also its neighbors.
If, for example, a static agent moves, the agent and all its neighbors will be
  affected, while a change to the agent's force threshold, which must be exceeded
  to move the agent, only affects itself.

\section{Evaluation}
\label{sec:evaluation}

\subsection{Benchmark Simulations}
\label{sec:benchmarks}

We use five simulations to evaluate the performance of the simulation engine:
  cell proliferation, cell clustering, and use cases in the domains of
  epidemiology, neuroscience, and oncology.
These simulations use double-precision floating point variables and are
  described in detail in \cite{breitwieser-bdm}.
Table~\ref{tab:sim-characteristics} shows that these simulations cover a broad
  spectrum of performance-related simulation characteristics and contain
  information about the number of agents, diffusion volumes, and iterations
  executed.
We set the number of agents between two and 12.6 million to keep the total
  execution time of all benchmarks manageable.
This is necessary due to the slow execution of the various baselines.
In addition, Section~\ref{sec:runtime-complexity} shows a benchmark in which
  each simulation is executed with one billion agents.
Also the comparison with Biocellion in Section~\ref{sec:eval:comparison-with-biocellion} 
  contains a benchmark with 1.72 billion cells.

\subsection{Experimental Setup and Reproducibility}

All tests were executed in a Docker container with an Ubuntu 20.04 based image.
Table~\ref{tab:systems} gives an overview of the main parameters of the three 
  servers we used to evaluate the performance of \bdm{}.
If it is not explicitly mentioned, assume that System~A was used to execute a benchmark.

We provide all code, the self-contained docker image, 
  more detailed 
  information on the hardware and software setup, and instructions to 
  execute the benchmarks in the supplementary materials 
  (\url{https://doi.org/10.5281/zenodo.6463816}).

\begin{table}
  \centering
  \footnotesize
  \caption{Performance-relevant simulation characteristics.}
\begin{tabular}{@{}rrrrrr@{}}
    \toprule
    \textbf{Characteristic} & 
    \rotatebox{90}{\textbf{Cell proliferation}} & 
    \rotatebox{90}{\textbf{Cell clustering}} & 
    \rotatebox{90}{\textbf{Epidemiology}} & 
    \rotatebox{90}{\textbf{Neuroscience}} & 
    \rotatebox{90}{\textbf{Oncology}} \\
    \midrule
    {Create new agents during simulation} &
    \xmark  &
      &
      &
    \xmark  &
    \xmark  \\

    {Delete agents during simulation} &
      &
      &
      &
      &
    \xmark  \\

    {Agents modify neighbors} &
      &
      &
      &
    \xmark  &
      \\

    {Load imbalance} &
      &
      &
    \xmark &
    \xmark &
      \\

    {Agents move randomly} &
      &
      &
    \xmark  &
      &
    \xmark  \\

    {Simulation uses diffusion} &
      &
    \xmark  &
      &
    \xmark  &
      \\

    {Simulation has static regions} & 
      &
      &
      &
    \xmark  &
      \\

    {Number of iterations} &
     500 &
     1000 &
     1000 &
     500 &
     288 \\
    
    {Number of agents (in millions)} &
     12.6 &
      2 &
     10 &
      9 &
     10 \\
    
    {Number of diffusion volumes} &
     0 &
     54m &
     0 &
     65k &
     0 \\

    \bottomrule
  \end{tabular}
\label{tab:sim-characteristics}
\end{table}

\begin{table}[htb]
\centering
 \caption{
   Benchmark hardware
}
 \resizebox{\linewidth}{!}{
 \begin{tabular}{@{}llp{7cm}l@{}}
   \toprule
   \textbf{System}
   & \textbf{Memory}
   & \textbf{CPU}
   & \textbf{OS} \\

   \midrule
   A
   & \vspace{0.5mm}504 GB\vspace{0.5mm} 
   & \multirow{2}{7cm}{\begin{tabular}[c]{@{}p{7cm}@{}}Four Intel(R) Xeon(R) E7-8890 v3 CPUs @ 2.50GHz with a total of 72 physical
   cores, two threads per core and four NUMA domains.\end{tabular}}
   & \multirow{2}{1cm}{\begin{tabular}[c]{@{}l@{}}\vspace{1mm}CentOS\\7.9.2009\end{tabular}} \\

   \cmidrule{1-2}
   B
   & \vspace{0.5mm}1008 GB\vspace{0.5mm} 
   &
   & \\

   \cmidrule{1-4}
   C
   & 62 GB
   & \begin{tabular}[c]{@{}p{7cm}@{}}Two Intel(R) Xeon(R) E5-2683 v3 CPUs @ 2.00GHz with a total of 28 physical cores, two threads per core and two NUMA domains.\end{tabular}
   & \begin{tabular}[c]{@{}l@{}}\vspace{1mm}CentOS\\Stream 8\end{tabular} \\

   \bottomrule
 \end{tabular}}
\label{tab:systems}
\end{table}

\subsection{General Performance Metrics}

We characterize the agent-based simulation workload by breaking down the
  operation's execution time, and exploring microarchitecture inefficiencies.

\begin{figure}[t]
  \centering
\begin{subfigure}{.49\linewidth}
\centering
    \includegraphics[width=\textwidth]{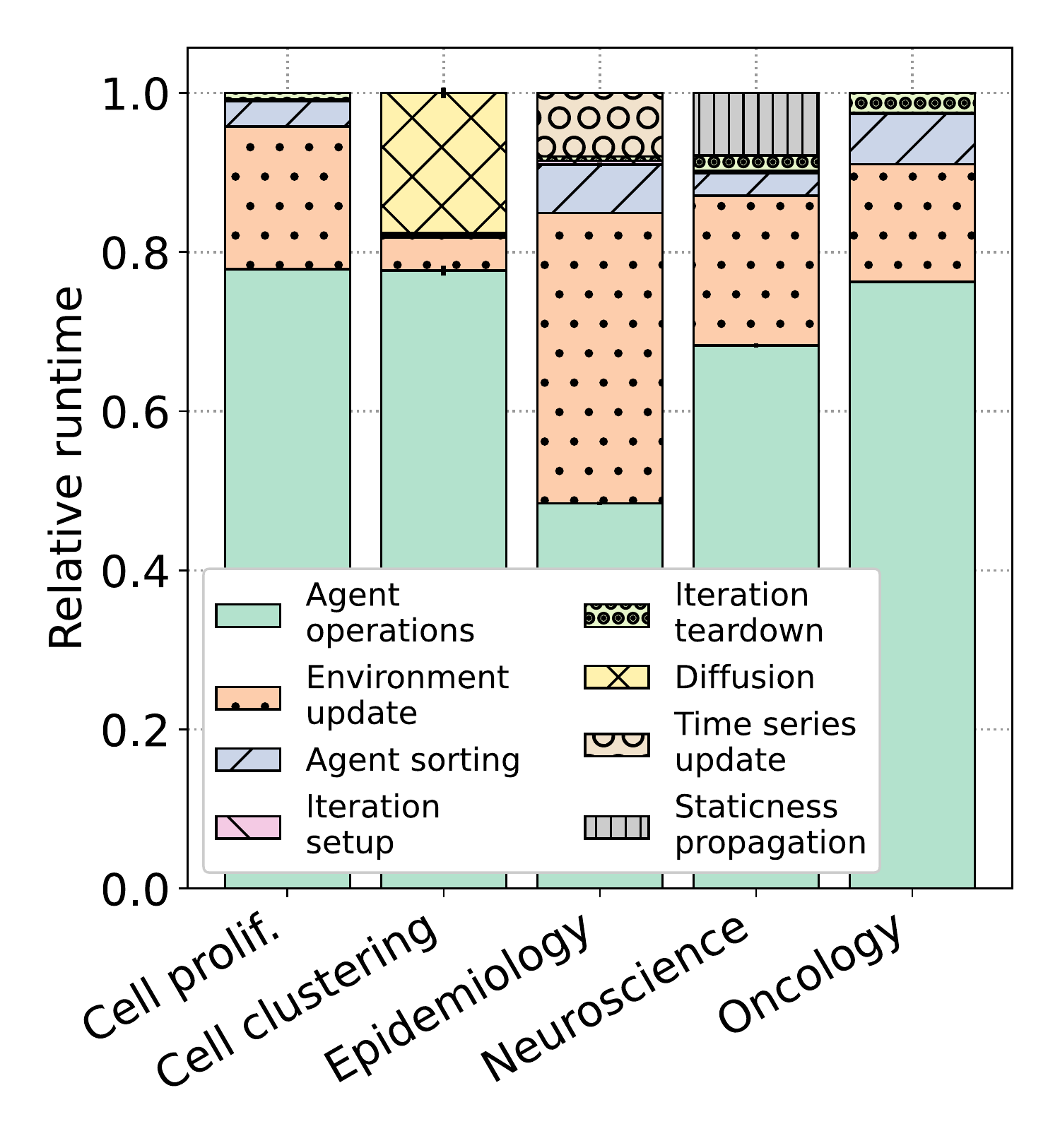}
\end{subfigure}
  \begin{subfigure}{.49\linewidth}
    \centering
    \includegraphics[width=\textwidth]{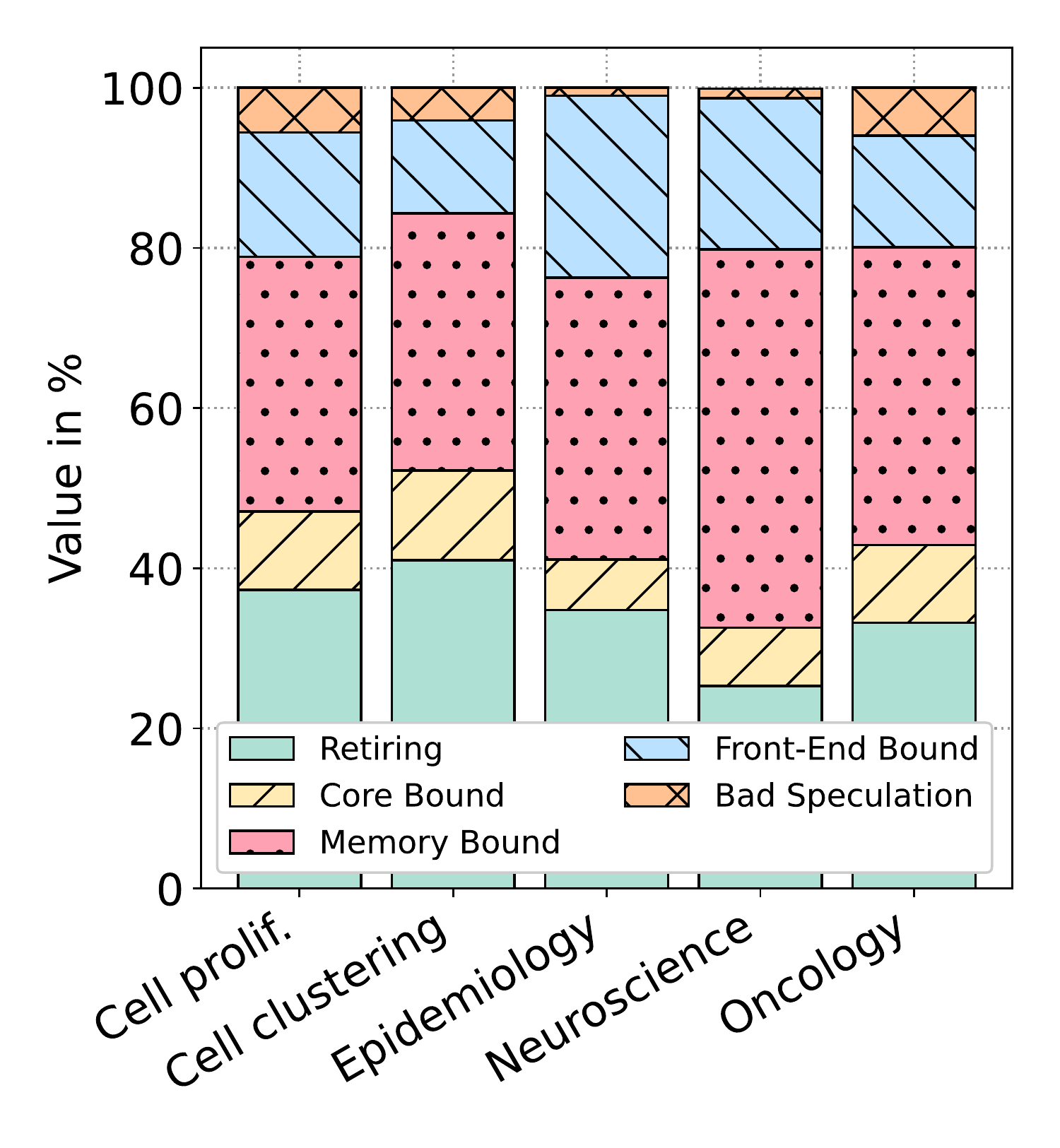}
\end{subfigure}
\caption{Operation runtime breakdown (left) and microarchitecture analysis
    (right)}
  \label{fig:operation-breakdown-uarch-analysis}
\end{figure}

The following benchmarks were performed with all optimizations enabled.
Figure~\ref{fig:operation-breakdown-uarch-analysis} shows a breakdown of all operations in our
  benchmark simulations.
 The majority of the runtime is spent in agent operations (median:
  \result{76.3\%}) which subsumes, among others, the execution of behaviors,
  calculation of mechanical forces, discretization, and detection of static
  regions.
Rebuilding the uniform grid environment at every time step is the second
  biggest runtime contributor, \result{4.09--36.5\%} (median: \result{18.0\%}).
The epidemiology use case considers a wider environment that manifests itself
  in an increased update time.
The average cost of agent sorting in its optimal setting (see
  Figure~\ref{fig:load-balancing}) is \result{0.180\%--6.33\%}.
Since adding and removing agents is parallelized, iterations' setup and tear
  down consume only \result{2.66\% (max)} of the execution time.

In the microarchitecture analysis, we observe that the benchmark simulations
  are primarily memory-bound.
We lose between \result{31.8 and 47.2\%} of processor pipeline slots because
  the operands are not available.

\subsection{Runtime and Space Complexity}
\label{sec:runtime-complexity}

\begin{figure}[tb]
  \centering
\begin{subfigure}{.49\linewidth}
    \centering
    \includegraphics[width=\textwidth]{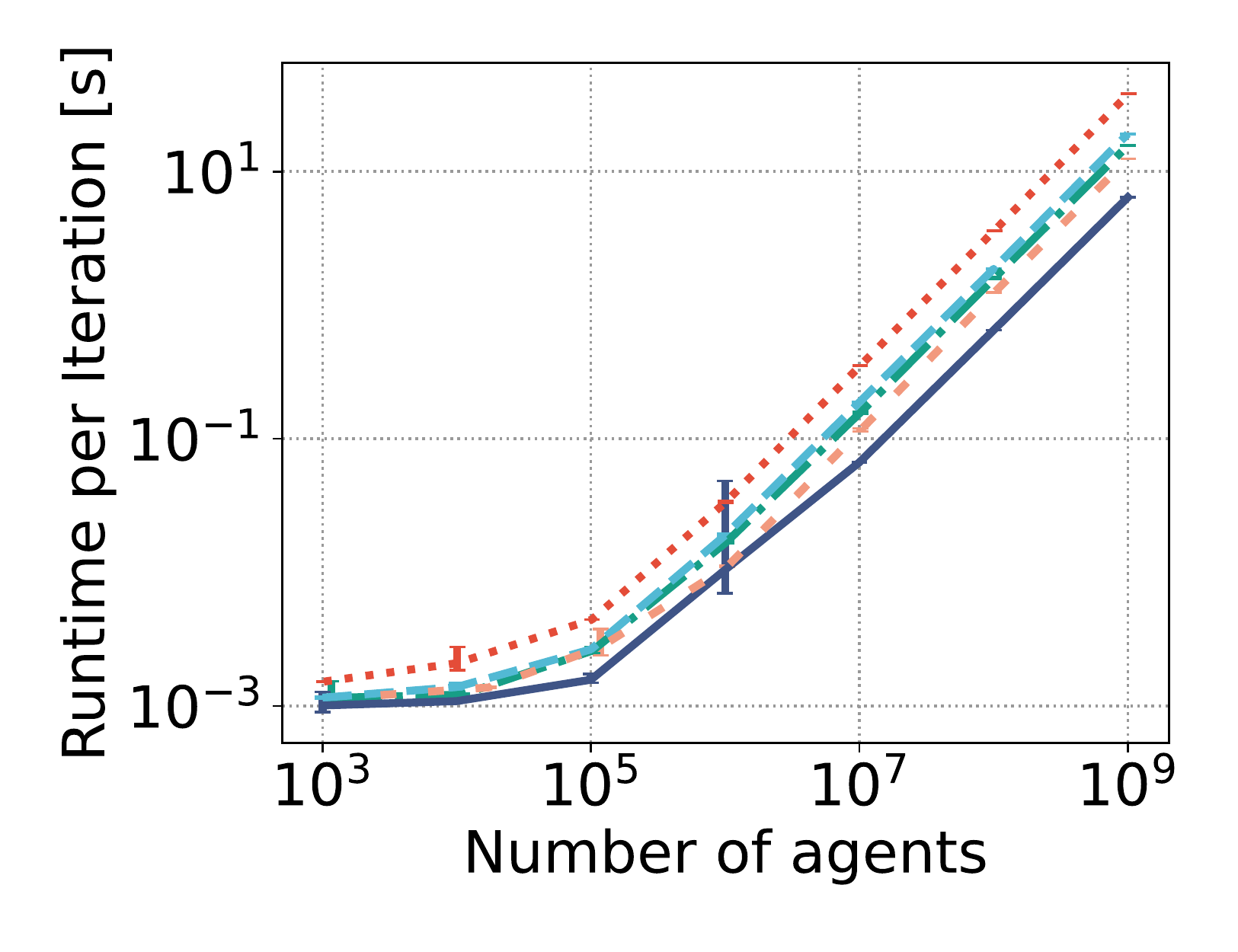}
\end{subfigure}
  \begin{subfigure}{.49\linewidth}
    \centering
    \includegraphics[width=\textwidth]{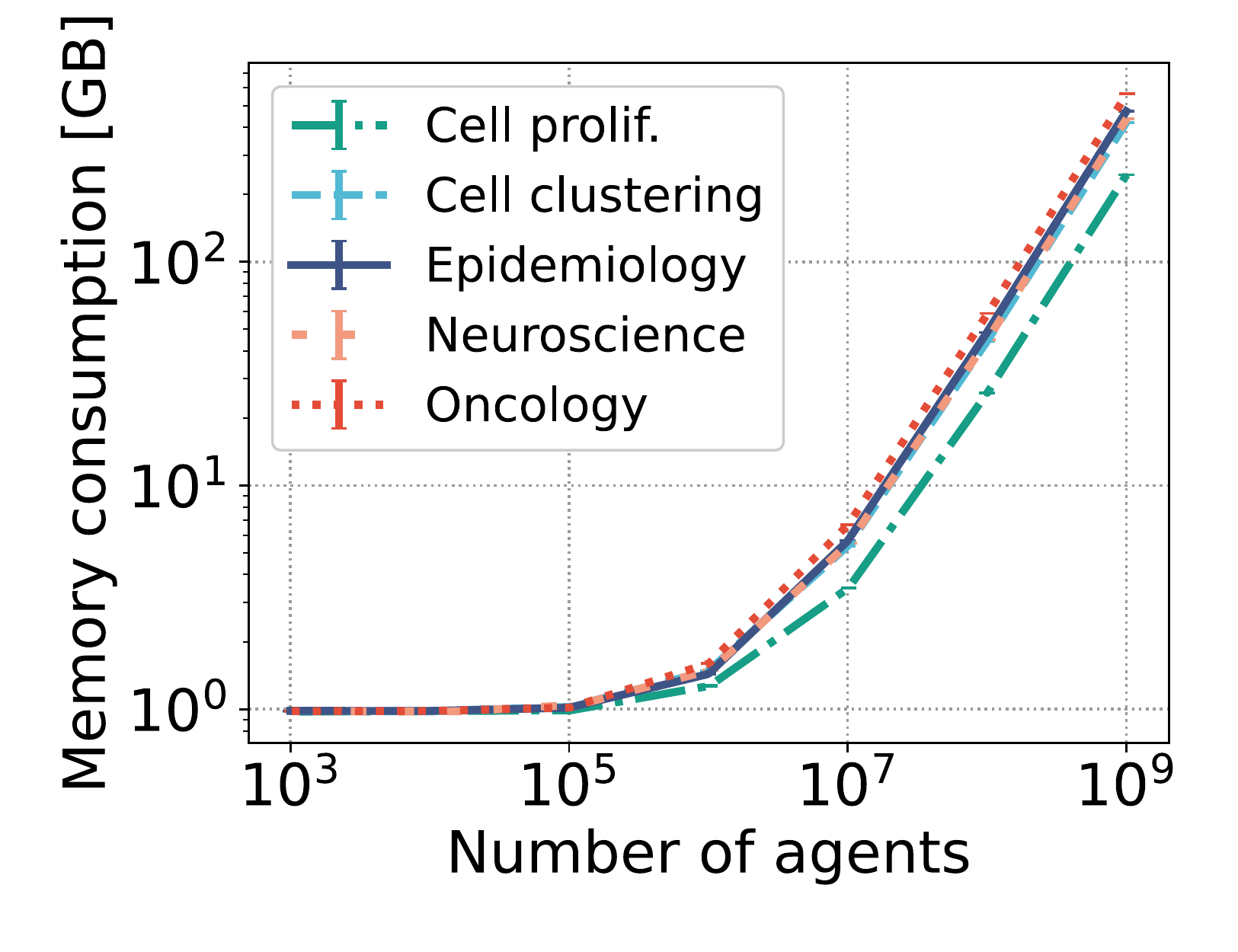}
\end{subfigure}
\caption{Average runtime per iteration and memory consumption analysis as
    the number of agents varies from $10^3$ to $10^9$}.
  \label{fig:runtime-complexity}
\end{figure}

We analyze the runtime and memory consumption of \bdm{} on System~B by increasing the
  number of agents from $10^3$ to $10^9$ for each simulation (Figure~\ref{fig:runtime-complexity}).
With one thousand agents, the execution time for one iteration is
  \result{on average 1.21 ms} and increases only slightly \result{until $10^5$ agents (2.80ms)}.
From there on, runtime increases \emph{linearly} to one billion agents in which
  one iteration takes between \result{6.41 and 38.1 seconds} to
  execute.
A similar trend can be observed for the memory consumption of \bdm{} (using
  double-precision floating point values), which remains below \result{$1.60$ GB
    until $10^6$} agents and increases \emph{linearly} to a maximum between
  \result{$245$ and $564$} GB.

The number of agents that \bdm{} can simulate is \emph{not} fundamentally
  limited to one billion.
The maximum depends only on the available memory of the underlying hardware and
  the tolerable execution time.

\subsection{Comparison with Biocellion}
\label{sec:eval:comparison-with-biocellion}

\begin{figure}[tb]
  \centering
  \begin{subfigure}{.25\linewidth}
    \includegraphics[width=\linewidth]{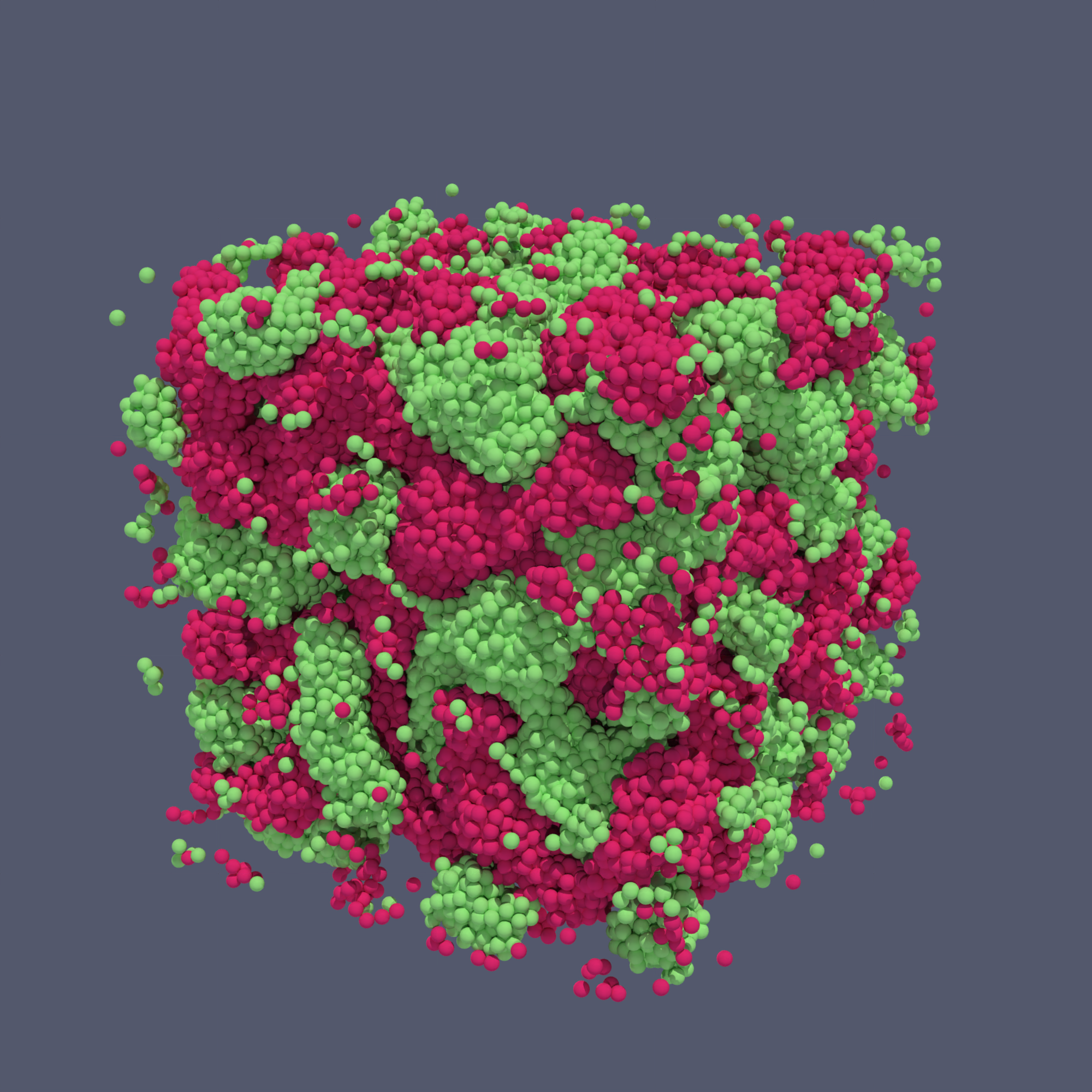}
    \caption{}
    \label{fig:biocellion-rendering}
  \end{subfigure}
  \begin{subfigure}{.72\linewidth}
    \includegraphics[width=0.325\linewidth]{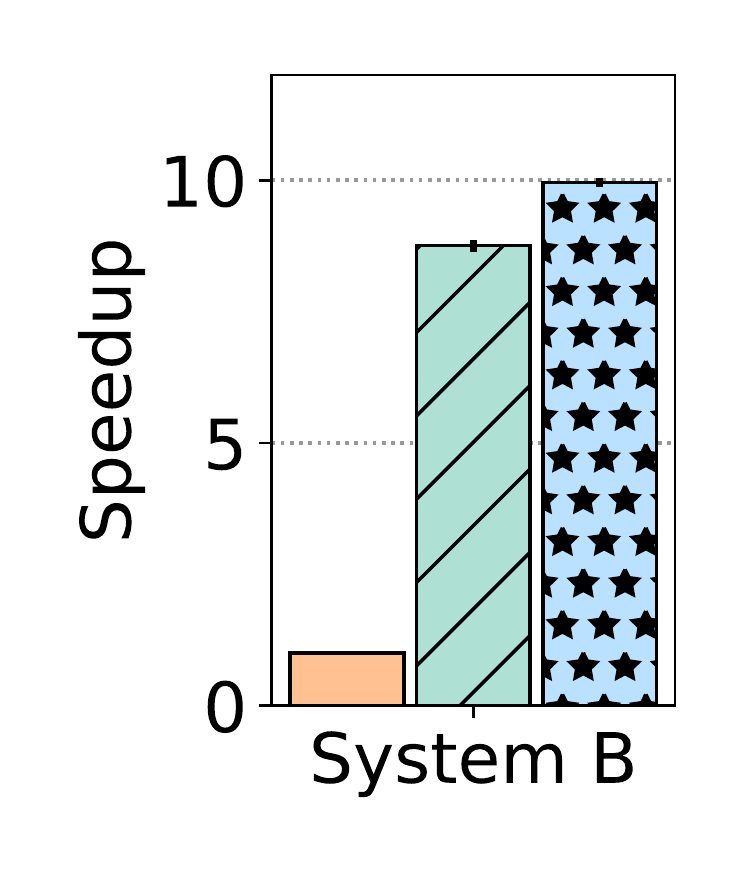}
    \includegraphics[width=0.655\linewidth]{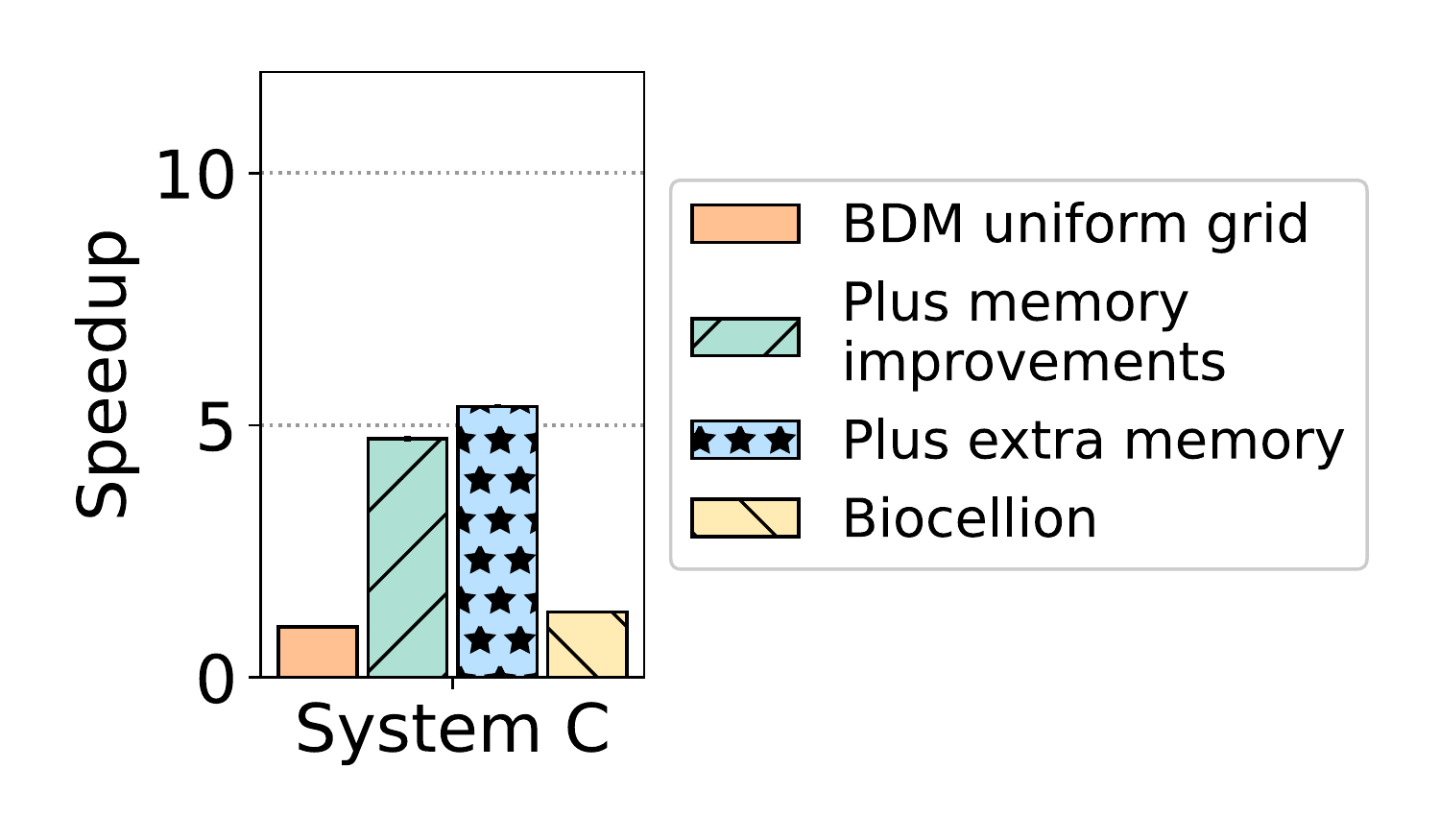}
    \caption{}
    \label{fig:biocellion-comparison}
  \end{subfigure}
    \caption{
      (a) Final simulation state after executing the Biocellion cell sorting model on \bdm{}.
      (b) Performance evaluation of the \bdm{} optimizations with a model with 28.6 million cells 
      on System~B (left) and System~C limited to 16 physical CPU cores (right).
      The Biocellion paper \cite{biocellion} provides only a performance measurement for the latter benchmark.
    }
  \label{fig:biocellion}
\end{figure}

We compare \bdm{} with Biocellion \cite{biocellion}, an agent-based framework for tissue models optimized
  for performance.
We implement the cell sorting simulation presented in the Biocellion paper (Section~3.1) in \bdm{}
  and use identical model parameters.
The visualization of the \bdm{} simulation with 50k cells (Figure~\ref{fig:biocellion-rendering})
  demonstrates a good agreement with the Biocellion results in Figure~3a in \cite{biocellion}.

Since we do not have access to the Biocellion code, because it is proprietary software, we compare \bdm{} to 
  the performance results provided in \cite{biocellion}.
First, we replicate the benchmark with 26.8 million agents using 16 CPU cores.
For Biocellion, Khang et al. \cite{biocellion} used a system with two Intel Xeon E5-2670 CPUs with 2.6 GHz.
We execute \bdm{} on System~C with a comparable CPU and limit the number of CPU cores to 16 to ensure a fair comparison. 
We observe that \bdm{} is \result{4.14$\times$} faster than Biocellion.
\bdm{} executes one iteration in 1.80s (averaged over 500 iterations), while Biocellion requires 7.48s.

Second, we consider the Biocellion benchmark in which Kang et al. executed 1.72 billion cells on a cluster with 4096 CPU cores 
  (128 nodes with two AMD Opteron 6271 Interlago 2.1 GHz CPUs per node). 
We execute the \bdm{} simulation with the same number of cells on a \emph{single} node (System~B). 
Although \bdm{} requires \result{26.3s} per iteration, which is \result{5.90$\times$} slower than Biocellion, \bdm{} uses 56.9$\times$ fewer CPU cores.
Therefore, we conclude that the performance per CPU core of \bdm{} is \result{9.64$\times$} more efficient than Biocellion.
We repeat the experiment with 281.4 million cells to verify the last observation.
Biocellion requires 4.37s per iteration (extracted from Figure~3b in \cite{biocellion}) using 21 nodes with a total of 672 CPU cores. 
The \bdm{} simulation on System~B with 72 CPU cores runs in almost identical \result{4.24s} per iteration.
This result confirms our observation that \bdm{} is an order of magnitude more efficient than Biocellion.

We evaluate the impact of our optimizations to provide insights into the question of why \bdm{} processes \result{4.14$\times$} more agents per CPU core in the first benchmark 
  and \result{9.64$\times$} in the second.
Therefore, we execute the relevant optimizations with 26.8 million cells on System~C limited to 16 CPU cores and System~B with 72 CPU cores.
Figure~\ref{fig:biocellion-comparison} shows that the difference can largely be explained by the memory optimizations having a more significant impact on machines with higher CPU core count.

\subsection{Comparison with Cortex3D and NetLogo}
\label{sec:eval:comparison-with-cx3d-netlogo}

\begin{figure}[t]
  \centering
    \includegraphics[width=\linewidth]{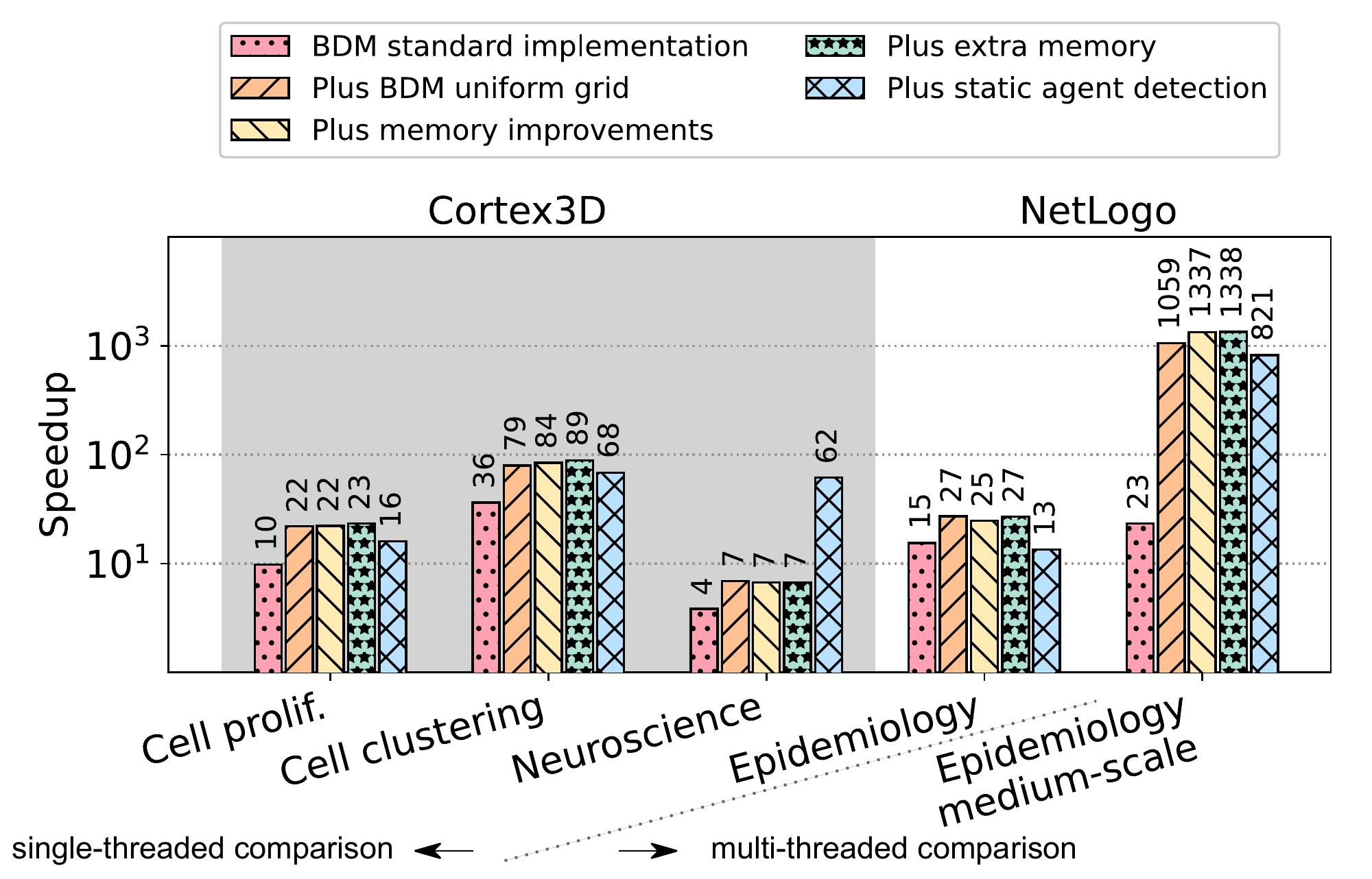}
\caption{Performance comparison with Cortex3D and NetLogo after the optimizations are progressively switched on.
}
  \label{fig:comparison-with-others}
\end{figure}

We also compare with capable single-thread tools to evaluate the parallel overhead of the \bdm{} implementation \cite{scalability_cost}.
We choose Cortex3D \cite{zublerdouglas2009framework} due to its similarity with the neuroscience 
  features of \bdm{} and select NetLogo \cite{netlogo} as a 
  representative for an easy-to-use general-purpose tool.
We extend the experiments from \cite{breitwieser-bdm} by analyzing the impact of the presented performance
  improvements and comparing the memory consumption.
This benchmark uses different simulation parameters for agents, diffusion
  volumes, and iterations, than shown in Table~\ref{tab:sim-characteristics}.
The first four benchmarks in Figure~\ref{fig:comparison-with-others} are
  small-scale benchmarks using between 2k and 30k agents and 0--128k diffusion
  volumes.
These benchmarks run for 100--1000 iterations and only use one thread because
  Cortex3D and NetLogo are not parallelized.
The ``epidemiology (medium-scale)'' benchmark contains 100k agents and uses 144
  threads.
NetLogo only benefits from parallel garbage collection in this scenario.
In the ``\bdm{} standard implementation'', all optimizations are turned off,
  and the kd-tree environment is used.

We make the following observations.
For the small-scale simulations using one thread, \bdm{} achieves a speedup of
  up to \result{78.8$\times$} while using \result{2.49$\times$} less memory.
We observe three orders of magnitude speedup and two orders of magnitude
  reduction in memory consumption for the medium-scale benchmark in which all
  threads were used.

The median speedup of the \bdm{} standard implementation is
  \result{15.5$\times$}.
The optimized uniform grid of \bdm{} boosts performance in all benchmarks
  (median: \result{2.18$\times$}) but has the most significant impact if
  parallelization is used (45.5$\times$).
Memory layout optimizations improve the runtime of medium-scale simulations by
  \result{26.2\%}, but not for small-scale ones.
The memory layout optimizations comprise the NUMA-aware iteration
  (Section~\ref{sec:numa-iteration}), agent sorting and balancing
  (Section~\ref{sec:load-balancing}), and memory allocator
  (Section~\ref{sec:mem-allocator}).
Due to the interdependency between these individual optimizations, we subsumed
  them into one category.
Similarly, extra memory usage during the agent sorting and balancing stage
  (Section~\ref{sec:load-balancing}) has only a slight performance impact (median
  speedup: \result{4.82\%}).
However, the static region optimization dramatically improves the performance
  in the neuroscience use case (speedup \result{9.22$\times$}).
Although the mechanism's overhead reduces the speedup for simulations without
  static regions, this is not problematic.
The modeler usually knows this characteristic a priori and only enables the
  mechanism if static regions are expected (see parameter
  \texttt{detect\_static\_agents}).

\subsection{Optimization Overview}

\begin{figure}[t]
  \centering
\begin{subfigure}{\linewidth}
    \centering
    \includegraphics[width=\linewidth]{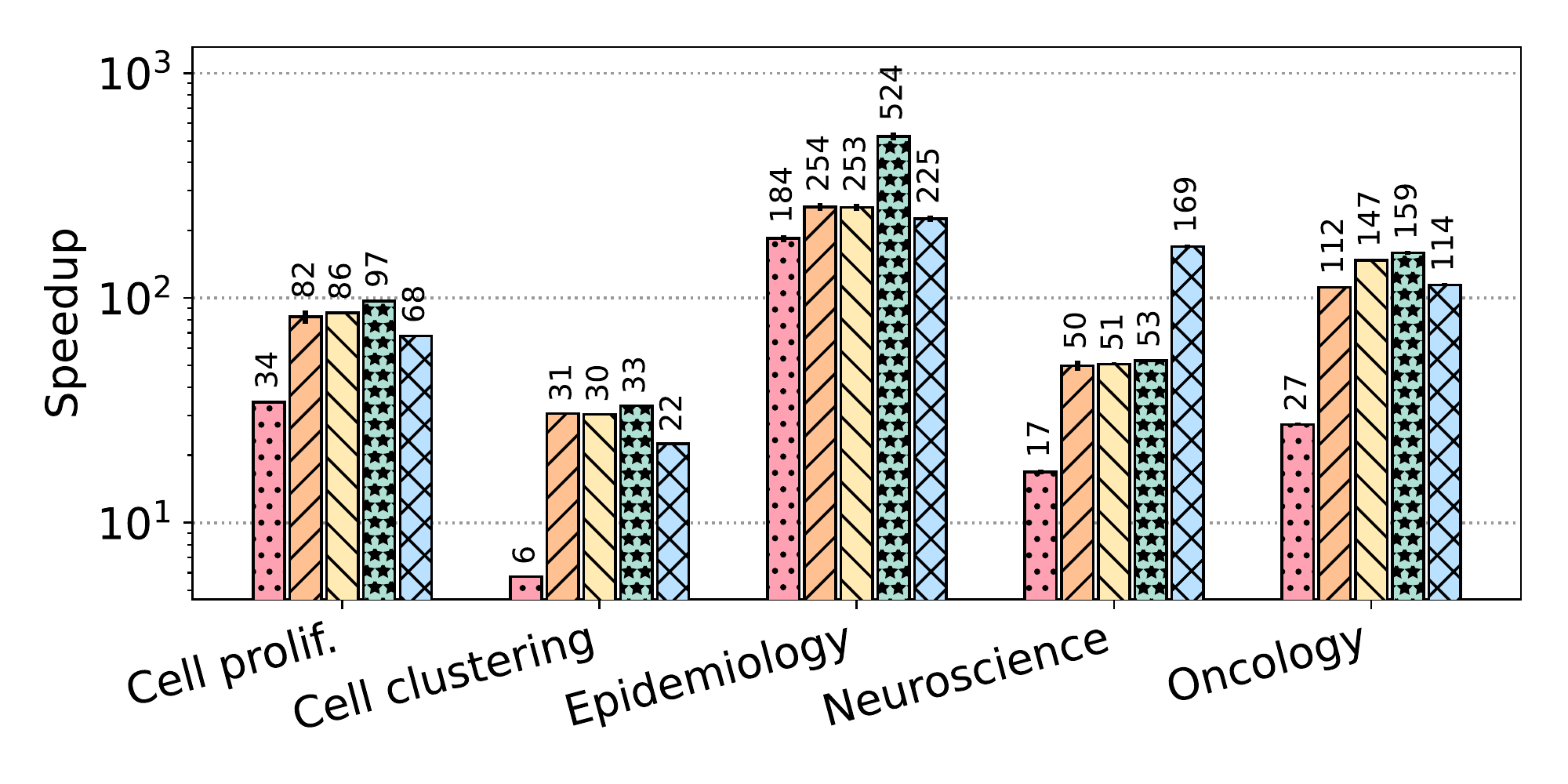}
\end{subfigure}
  \begin{subfigure}{\linewidth}
    \centering
    \includegraphics[width=\linewidth]{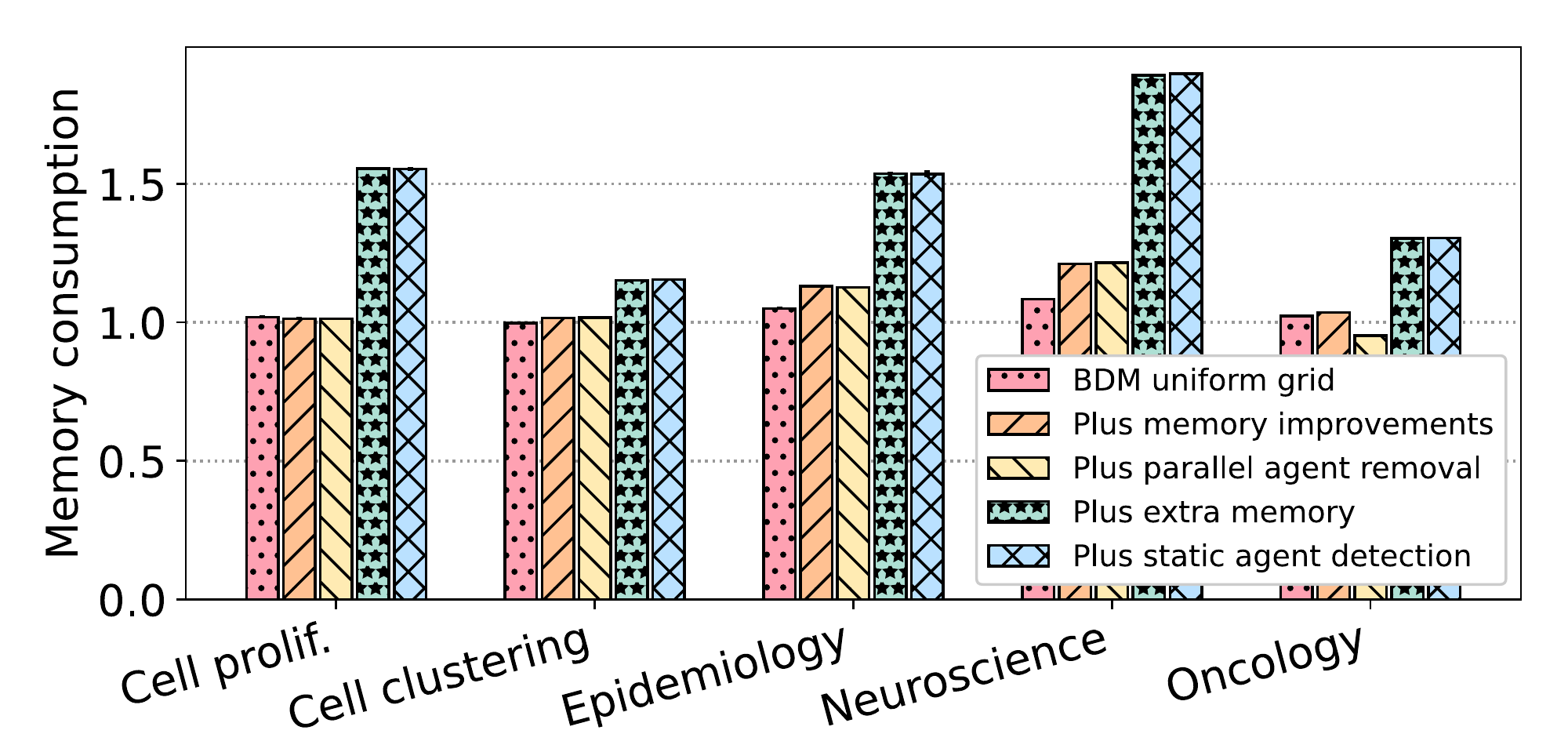}
\end{subfigure}
\caption{Speedup (top) and memory consumption (bottom)
    compared with the \bdm{} standard implementation after the
    optimizations are progressively switched on.
    The legend is shared between the plots.
  }
  \label{fig:optimization-overview}
\end{figure}

We assess the performance of the presented optimizations using larger-scale
  simulations (Table~\ref{tab:sim-characteristics}) by enabling optimizations
  step-by-step (Figure~\ref{fig:optimization-overview}).
The baseline in this comparison is the \bdm{} standard implementation
  introduced in Section~\ref{sec:eval:comparison-with-cx3d-netlogo}.

We make the following observations.
The \bdm{} optimizations improve overall performance between
  \result{33.1$\times$} and \result{524$\times$} (median: \result{159$\times$}).
These benchmarks confirm the speedup of \bdm{}'s optimized uniform grid that we
  observed in comparison with Cortex3D and NetLogo.
For these larger-scale simulations, the magnitude of the speedup increases up
  to \result{184$\times$} with a median of \result{27.4$\times$}.
A similar observation can be made for the static region detection mechanism,
  albeit with reduced magnitude (speedup: \result{3.22$\times$}).
The main difference between the comparison with Cortex3D and NetLogo and this benchmark
  is the impact of the memory layout optimizations of agents and behaviors and
  the usage of extra memory during agent sorting.
The maximum speedup is up to \result{5.30$\times$} (median:
  \result{2.96$\times$}) and up to \result{2.07$\times$} (median:
  \result{1.09$\times$}), respectively.
Only the Biocellion benchmark in Figure~\ref{fig:biocellion-comparison} shows a bigger impact.

The simulation time of the oncology use case, the only benchmark that removes
  agents from the simulation is reduced by \result{31.7\%} using the ``parallel
  removal'' optimization described in Section~\ref{sec:parallel-remove}.
The optimizations increase the median memory consumption by a mere
  \result{1.77\%}, which increases to \result{55.6\%} by enabling the use of extra
  memory during agent sorting.

\subsection{Scalability}

We evaluate the scalability of \bdm{} using the complete simulations lasting
  between 288 and 1000 iterations and perform a strong scaling analysis with
  different optimizations enabled.
The strong scaling analysis is performed with ten iterations.

\begin{figure}[htb!]
  \centering
\begin{subfigure}{.49\linewidth}
    \centering
    \includegraphics[width=0.95\textwidth]{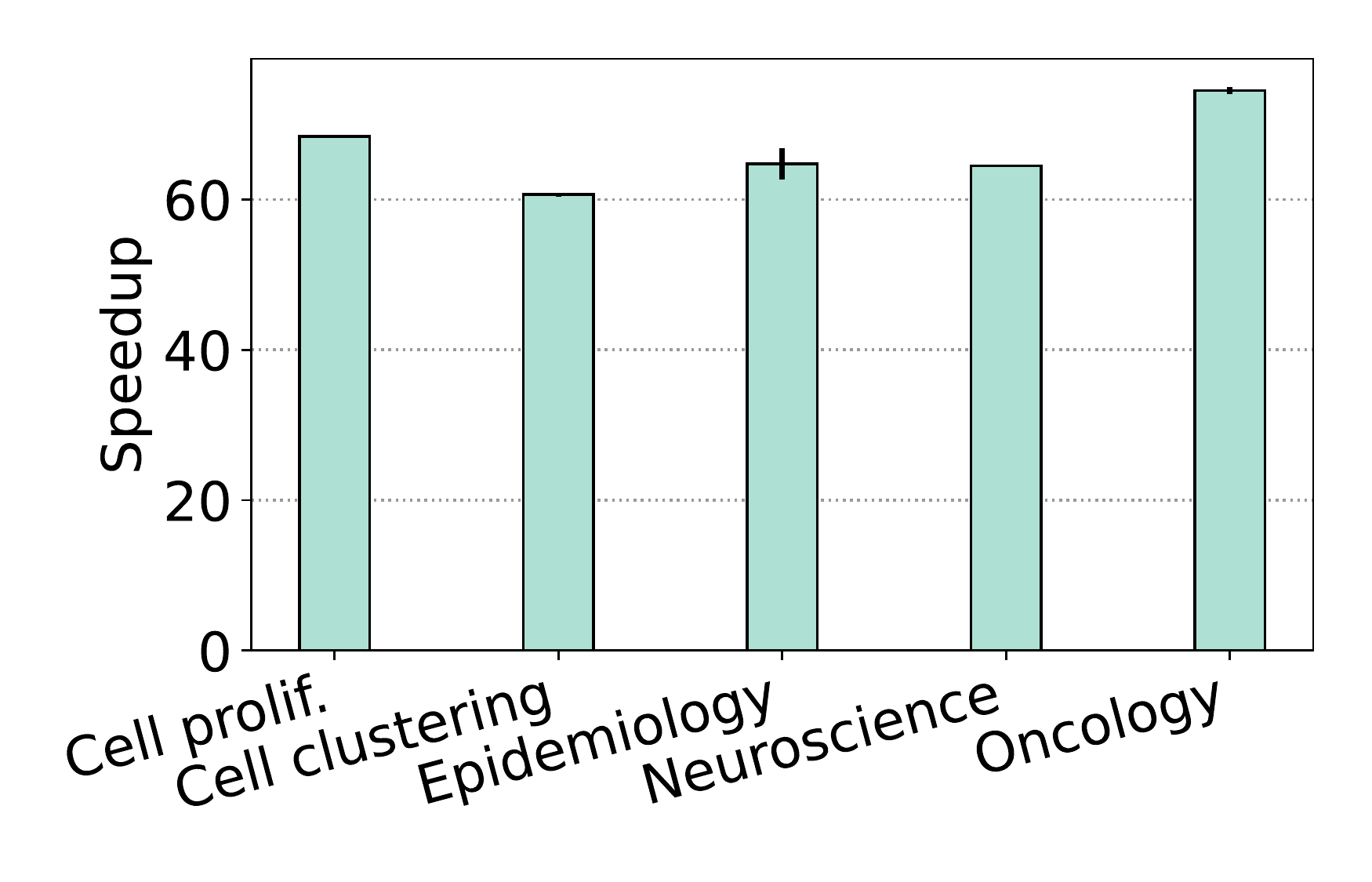}
    \caption{Whole simulation scalability}
    \label{fig:full-scalability}
  \end{subfigure}
  \begin{subfigure}{.49\linewidth}
    \centering
    \vspace{4mm}
    \includegraphics[width=0.9\textwidth]{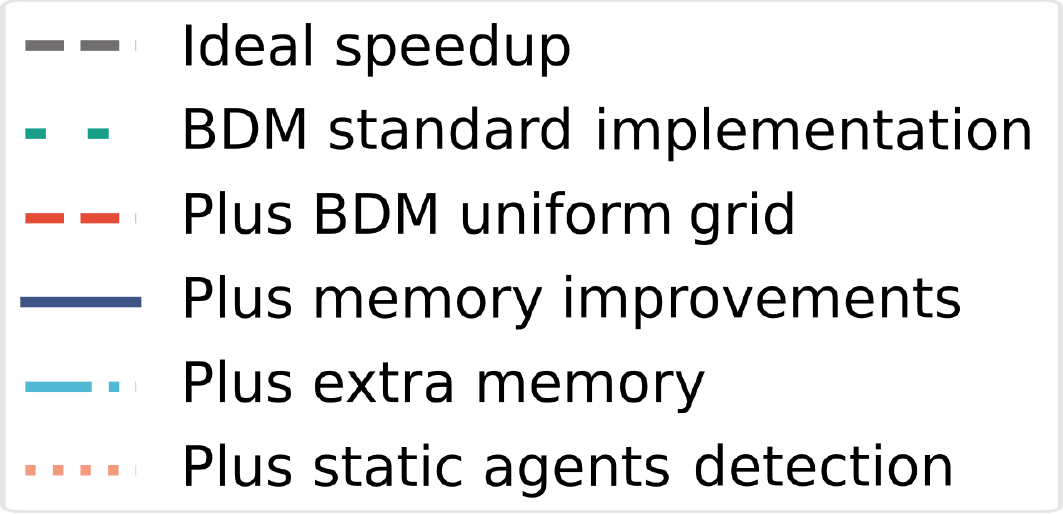}
    \vspace{4mm}
    \caption{Legend for (c)--(g)}
  \end{subfigure}
  \begin{subfigure}{\linewidth}
    \centering
    \vspace{2mm}
    \includegraphics[width=0.49\textwidth,trim=6mm 8mm 8mm 8mm, clip]{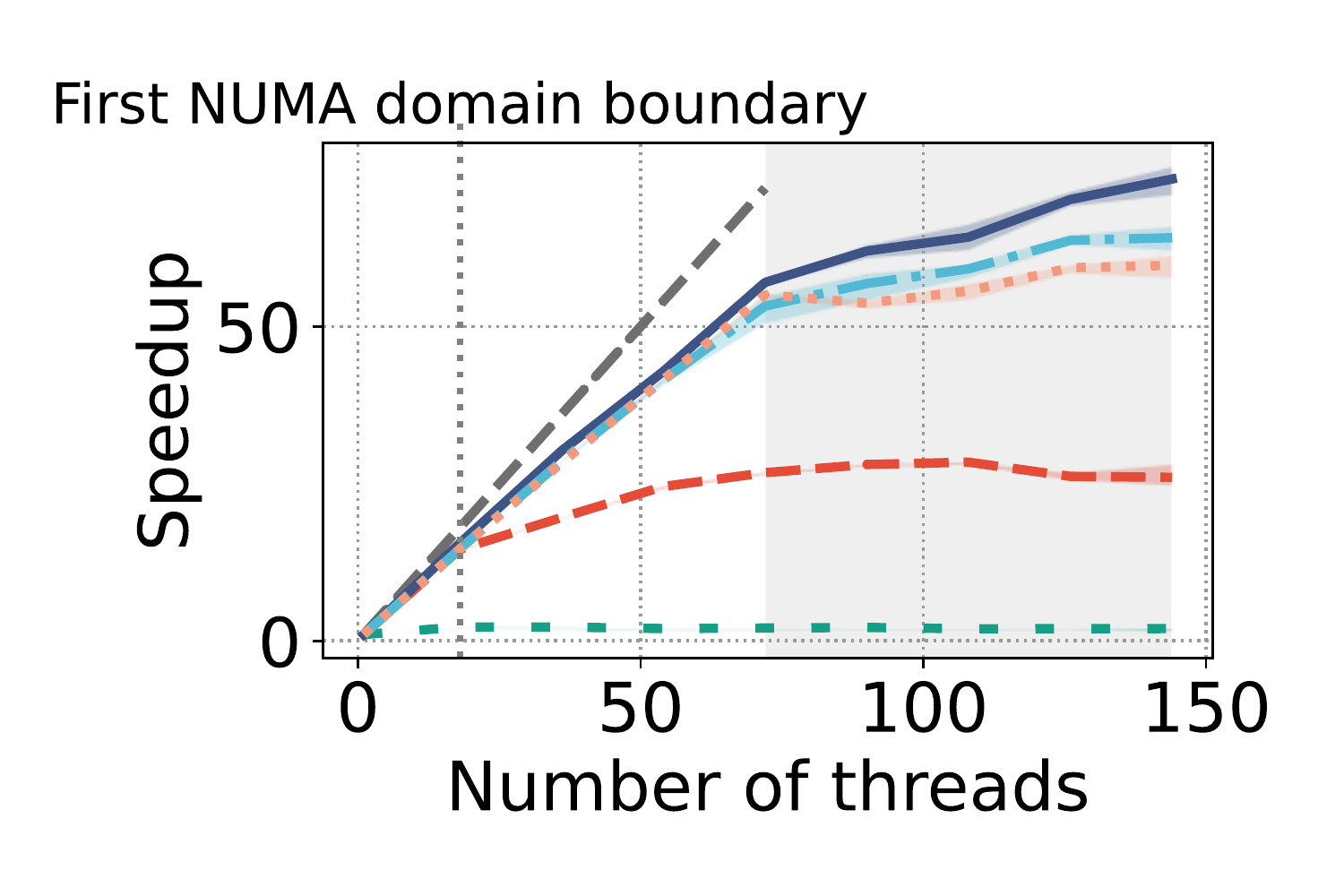}
    \includegraphics[width=0.49\textwidth,trim=6mm 8mm 8mm 8mm, clip]{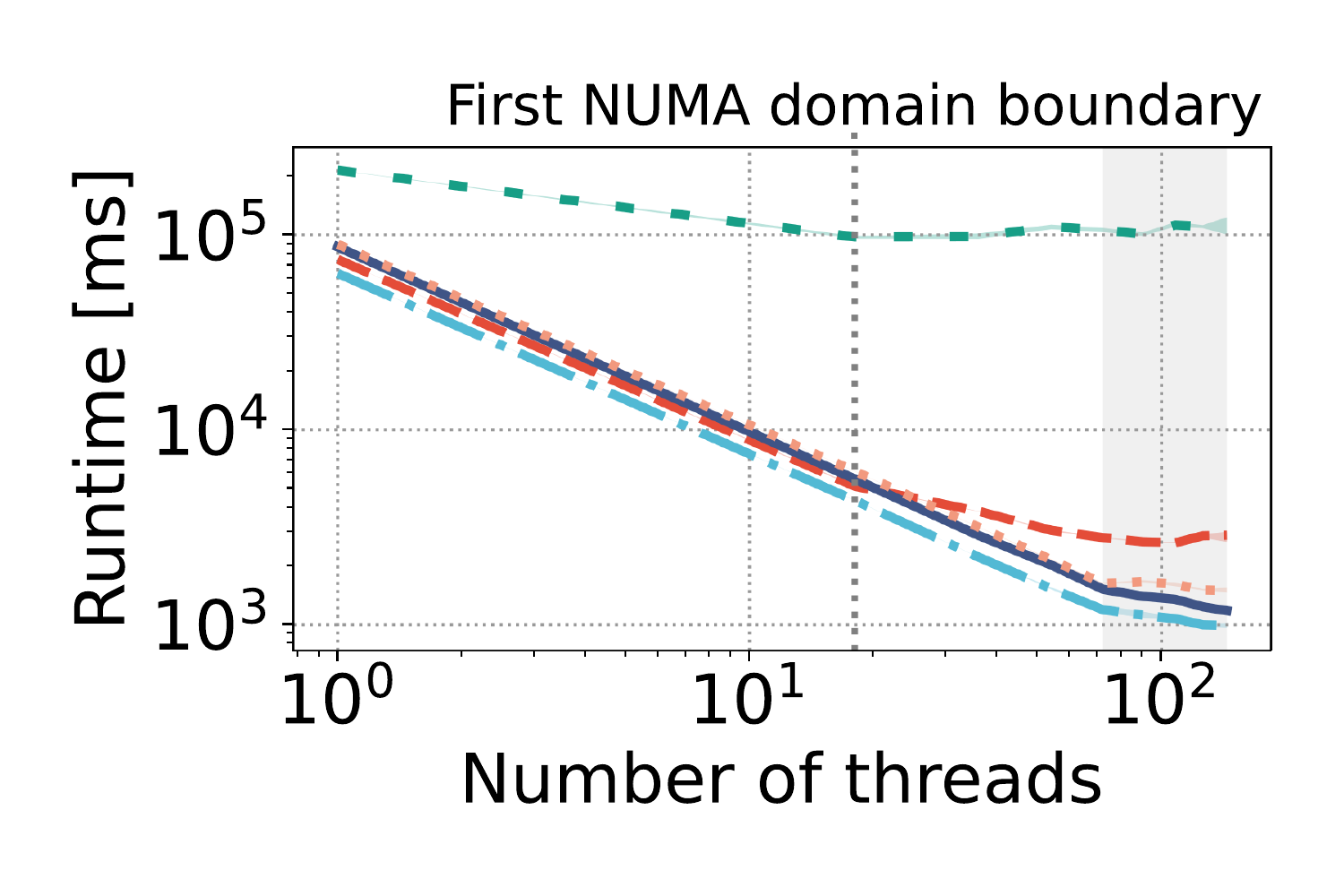}
    \caption{Cell proliferation}
    \label{fig:scalability-cgd}
  \end{subfigure}
  \begin{subfigure}{\linewidth}
    \centering
    \includegraphics[width=0.49\textwidth,trim=6mm 8mm 8mm 8mm, clip]{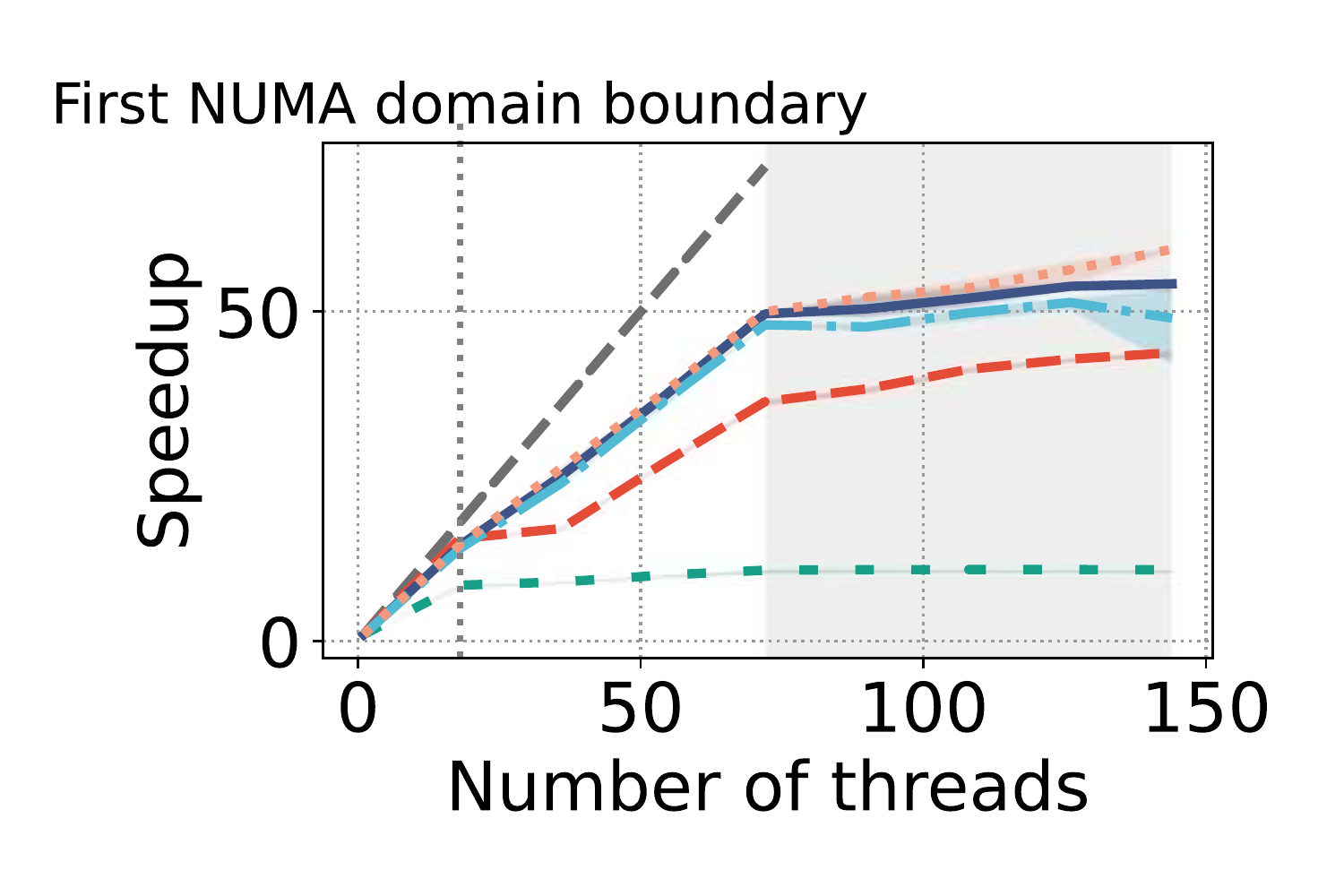}
    \includegraphics[width=0.49\textwidth,trim=6mm 8mm 8mm 8mm, clip]{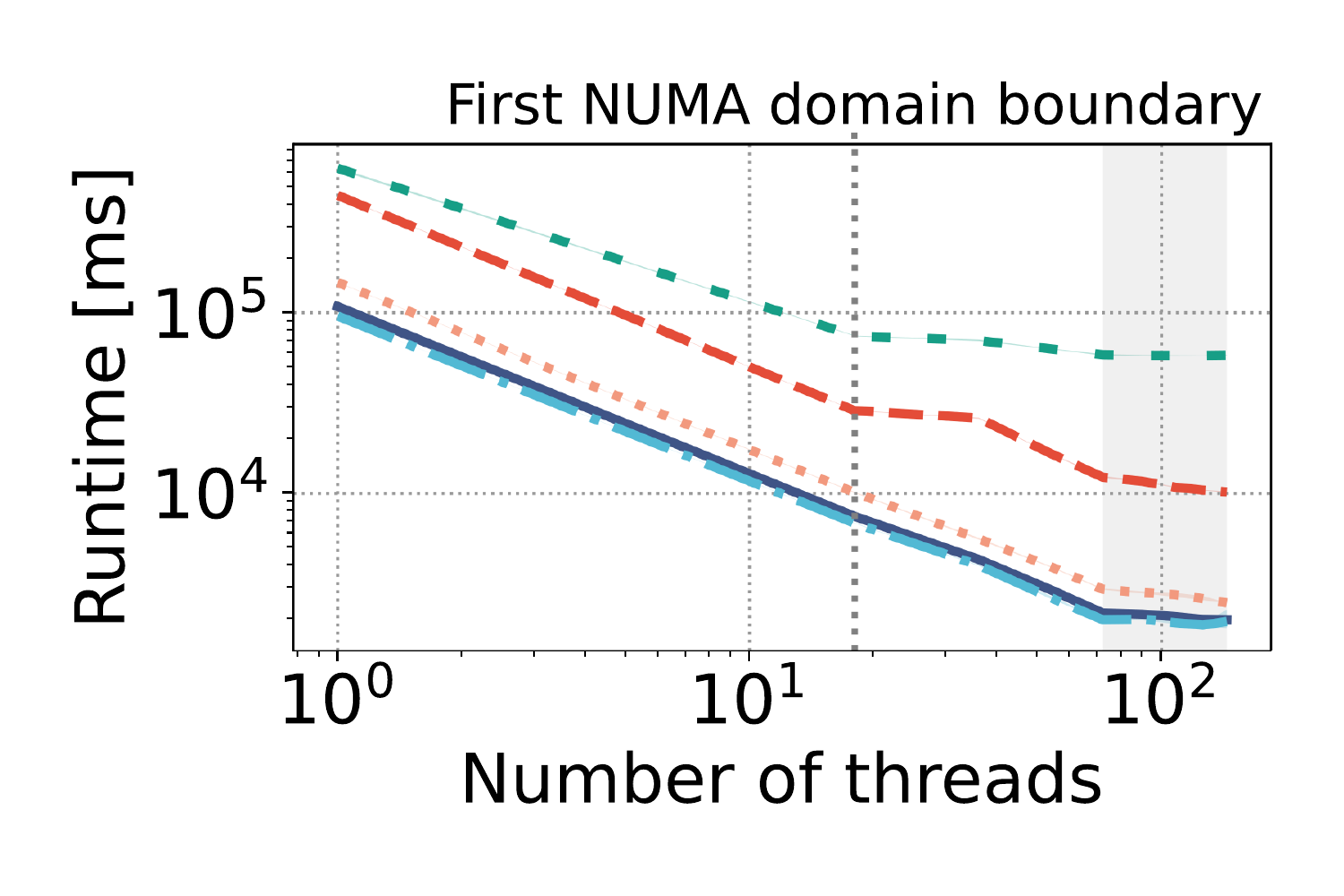}
    \caption{Cell clustering}
  \end{subfigure}
  \begin{subfigure}{\linewidth}
    \centering
    \includegraphics[width=0.49\textwidth,trim=6mm 8mm 8mm 8mm, clip]{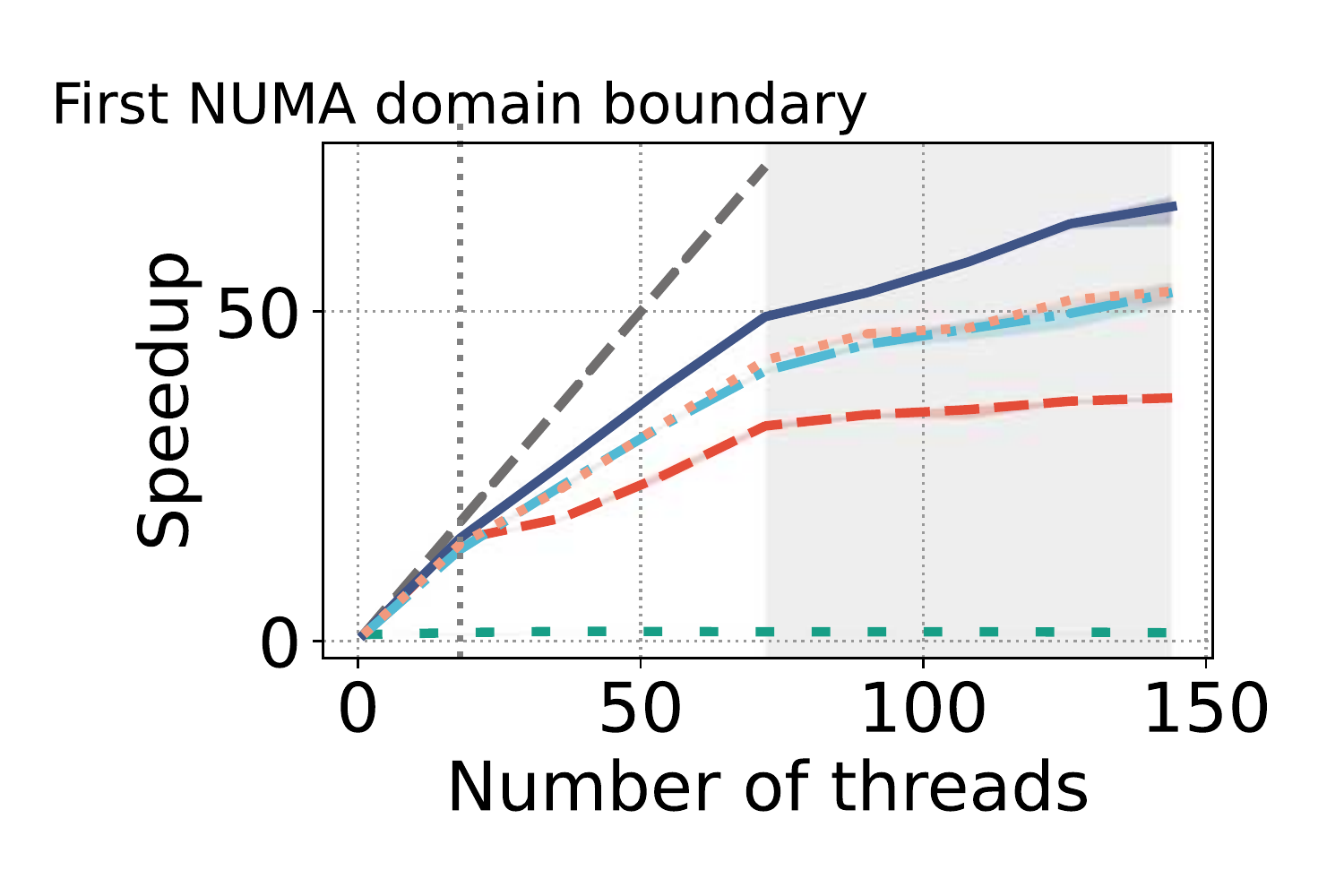}
    \includegraphics[width=0.49\textwidth,trim=6mm 8mm 8mm 8mm, clip]{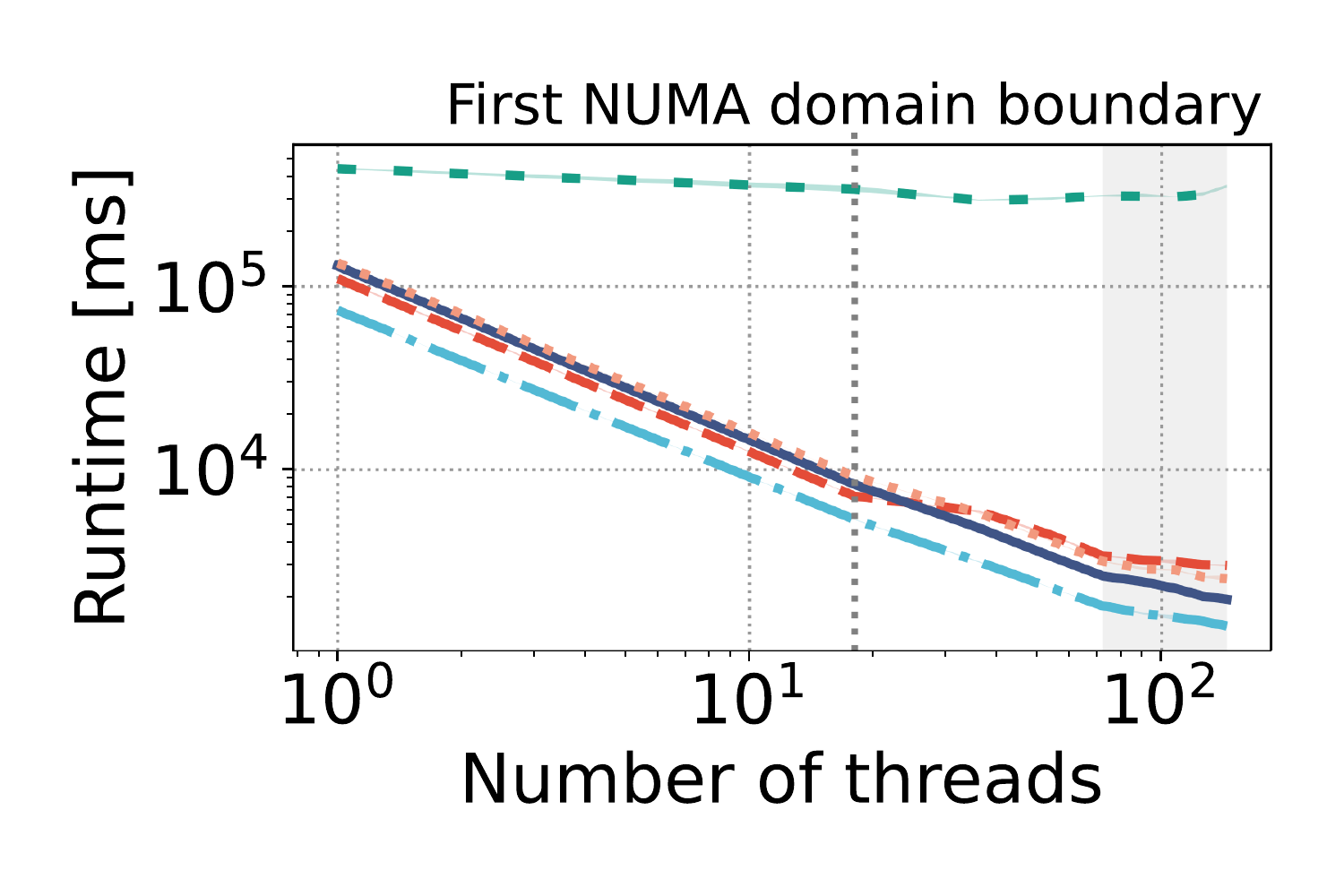}
    \caption{Epidemiology}
  \end{subfigure}
  \begin{subfigure}{\linewidth}
    \centering
    \includegraphics[width=0.49\textwidth,trim=6mm 8mm 8mm 8mm, clip]{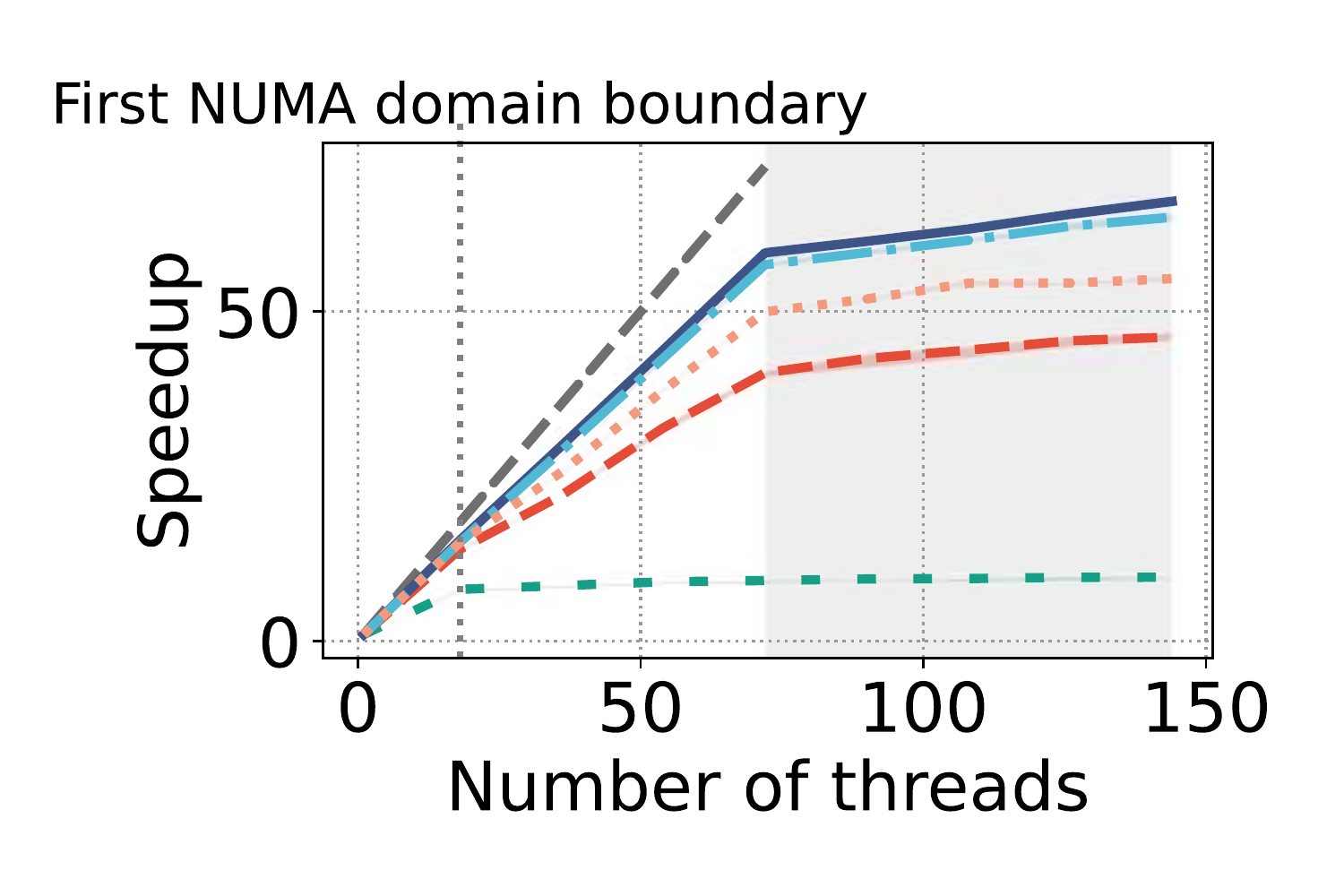}
    \includegraphics[width=0.49\textwidth,trim=6mm 8mm 8mm 8mm, clip]{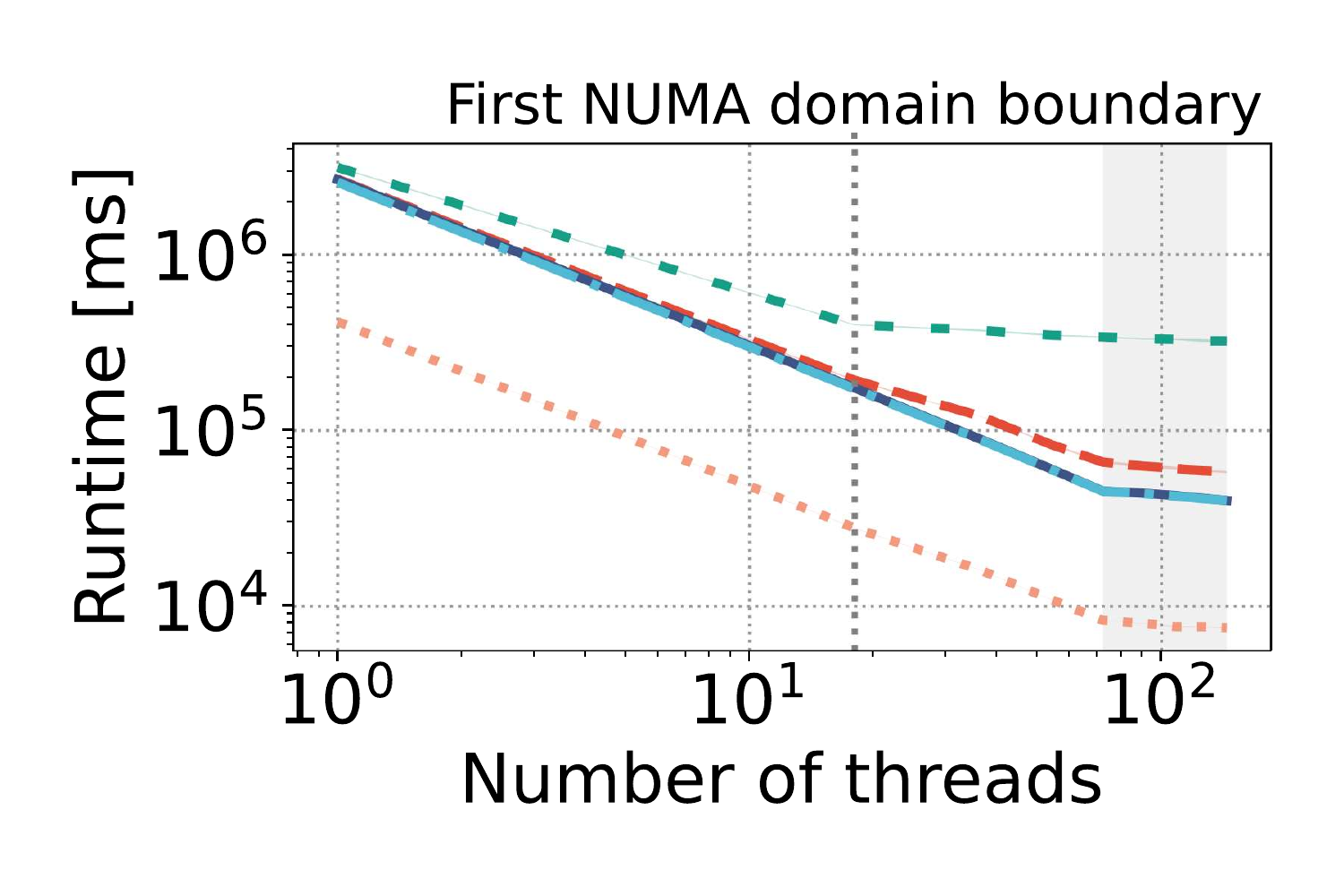}
    \caption{Neuroscience}
  \end{subfigure}
  \begin{subfigure}{\linewidth}
    \centering
    \includegraphics[width=0.49\textwidth,trim=6mm 8mm 8mm 8mm, clip]{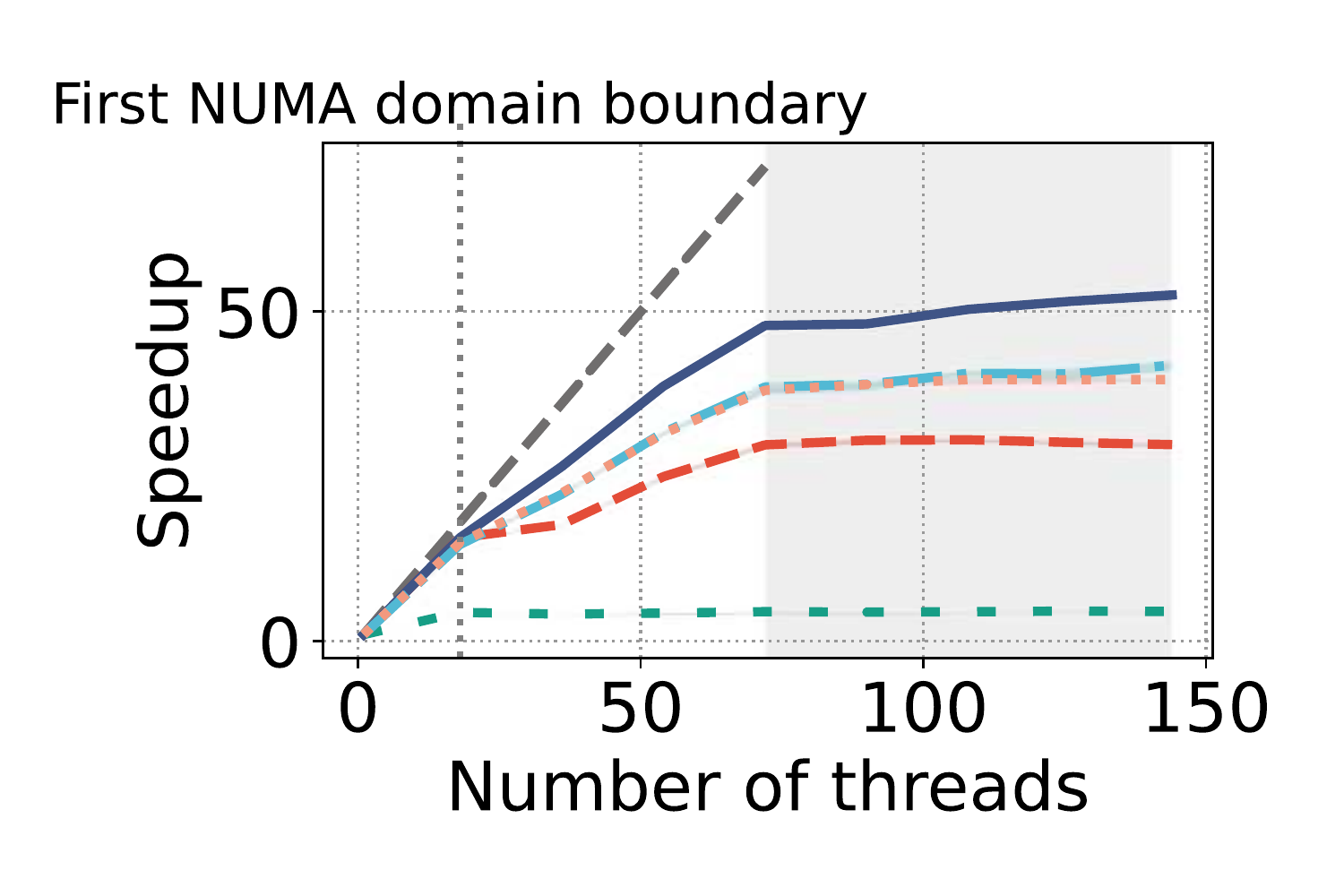}
    \includegraphics[width=0.49\textwidth,trim=6mm 8mm 8mm 8mm, clip]{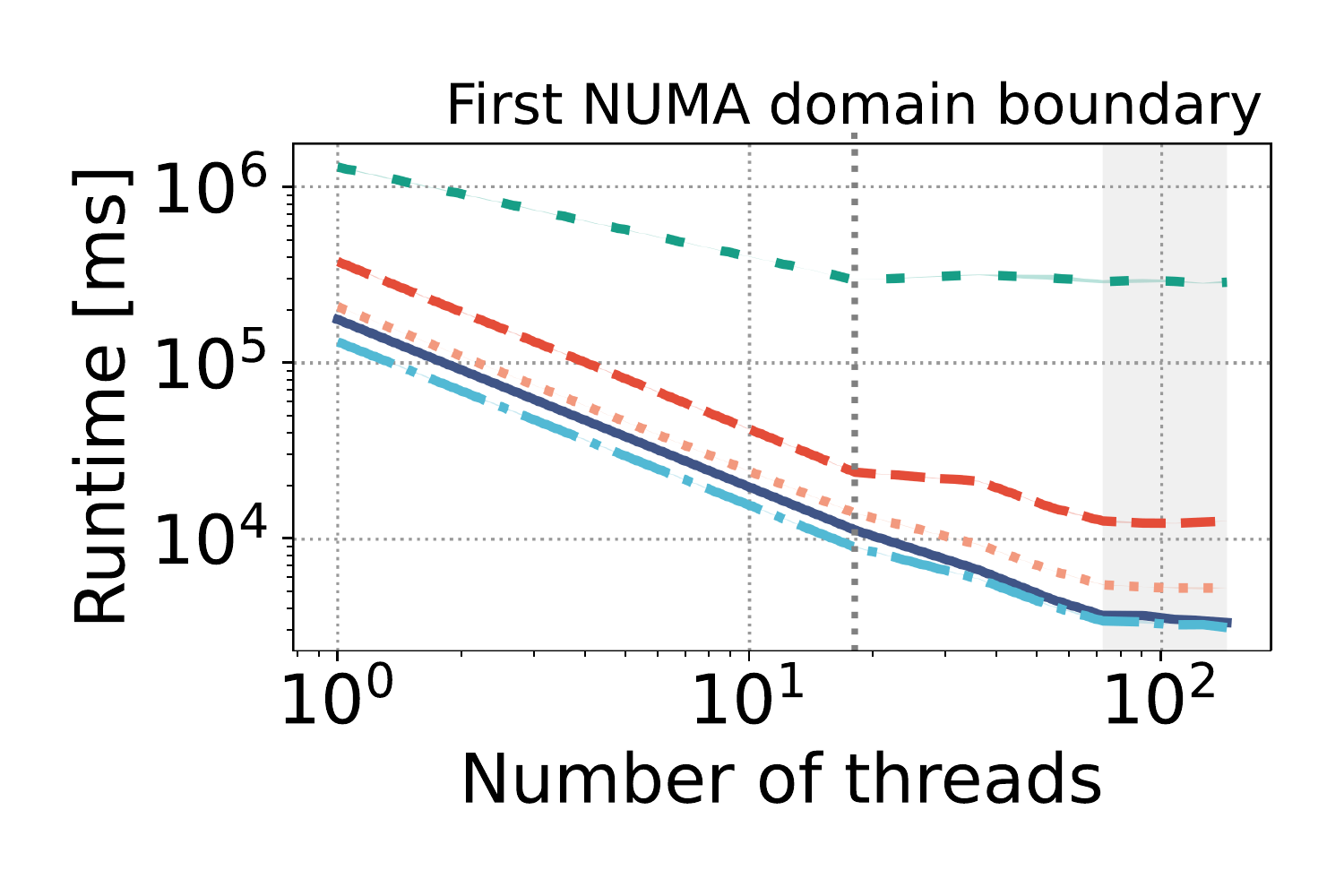}
    \caption{Oncology}
    \label{fig:scalability-ts}
  \end{subfigure}
\caption{
    (a) Simulation scalability using the whole simulation. 
    (b--g) Detailed strong scaling analysis using only ten time steps.
    The left column shows the speedup with respect to a single-thread execution, while the right column presents
    the total runtime.
}
\label{fig:scalability}
\end{figure}

Figure~\ref{fig:full-scalability} illustrates the excellent scalability of
  \bdm{} for complete simulations (i.e., executing all iterations).
The speedup using 72 physical cores with hyperthreading enabled is between
  \result{60.7$\times$ and 74.0$\times$ (median 64.7$\times$)} compared to serial
  execution.
Section~\ref{sec:eval:comparison-with-cx3d-netlogo} shows that \bdm{} with one CPU
  core is more than \result{23$\times$} faster than Cortex3D.
If we combine this result with the scalability analysis, which shows that
  \bdm{} with 72 CPU cores is more than \result{60$\times$} faster than one CPU
  core, we can conclude that \bdm{} is up to three orders of magnitude faster
  than Cortex3D.

Figures~\ref{fig:scalability-cgd}--\ref{fig:scalability-ts} show the strong scaling analysis 
  for each benchmark simulation with ten iterations after progressively switching on the 
  presented optimizations. 
The left column shows the speedup with respect to a single-thread execution, and the 
  right column presents the average runtime in milliseconds to highlight the absolute differences 
  between various optimizations and the reduction in runtime with increasing threads.
We make the following observations.
The \bdm{} standard implementation scales poorly due to the serial build of the
  kd-tree environment, which is improved considerably by using \bdm{}'s optimized
  uniform grid (Section~\ref{sec:grid}).
The presented memory optimizations (Section~\ref{sec:opt:memory-layout}) fully
  achieve their desired effect and allow \bdm{} to scale across NUMA domains and
  high CPU-core counts.

\subsection{Neighbor Search Algorithm Comparison}
\label{sec:eval:environment}

\begin{figure}[t]
  \centering
\begin{subfigure}{\linewidth}
    \centering
    \includegraphics[width=0.49\textwidth]{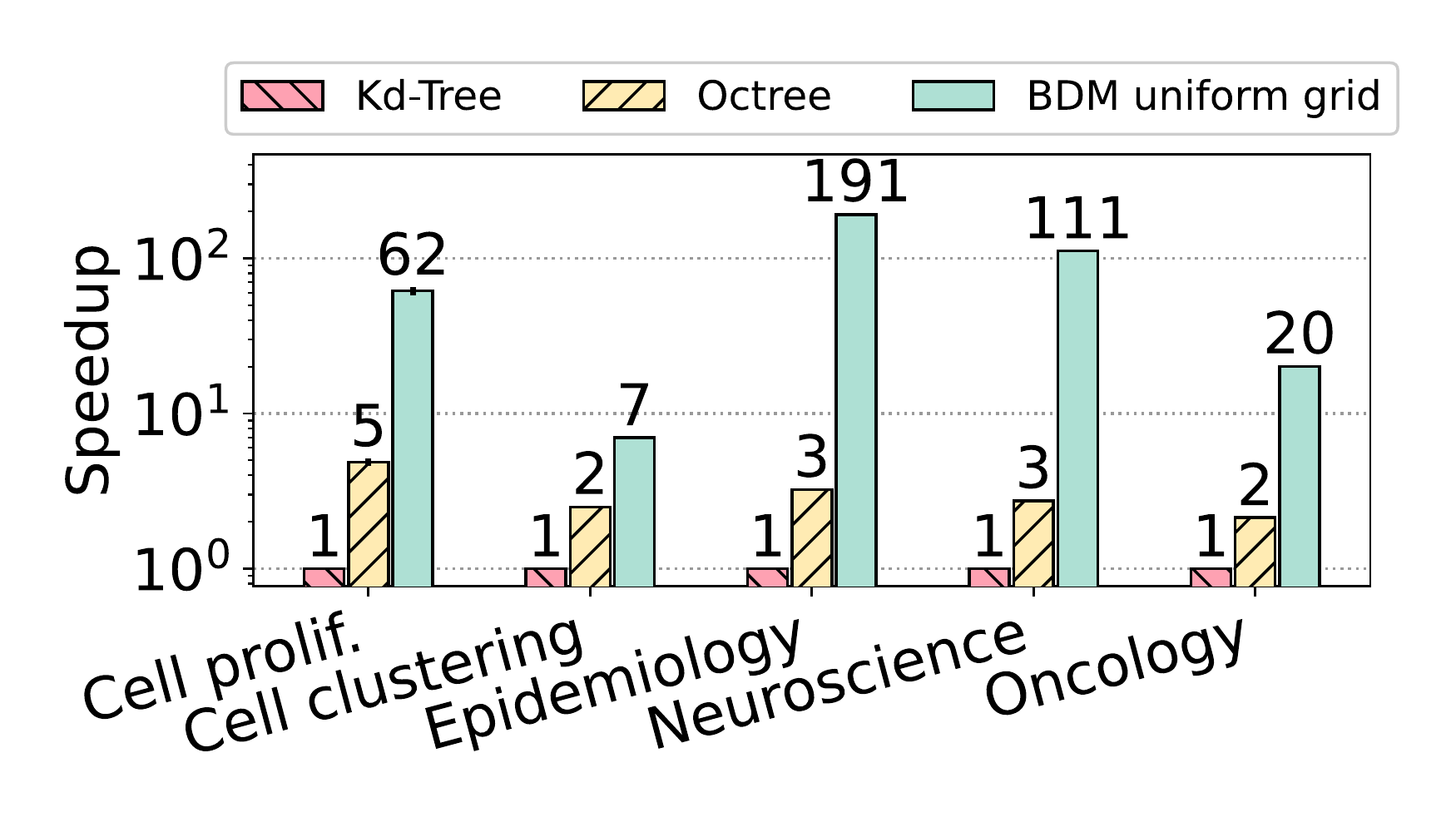}
    \includegraphics[width=0.49\textwidth]{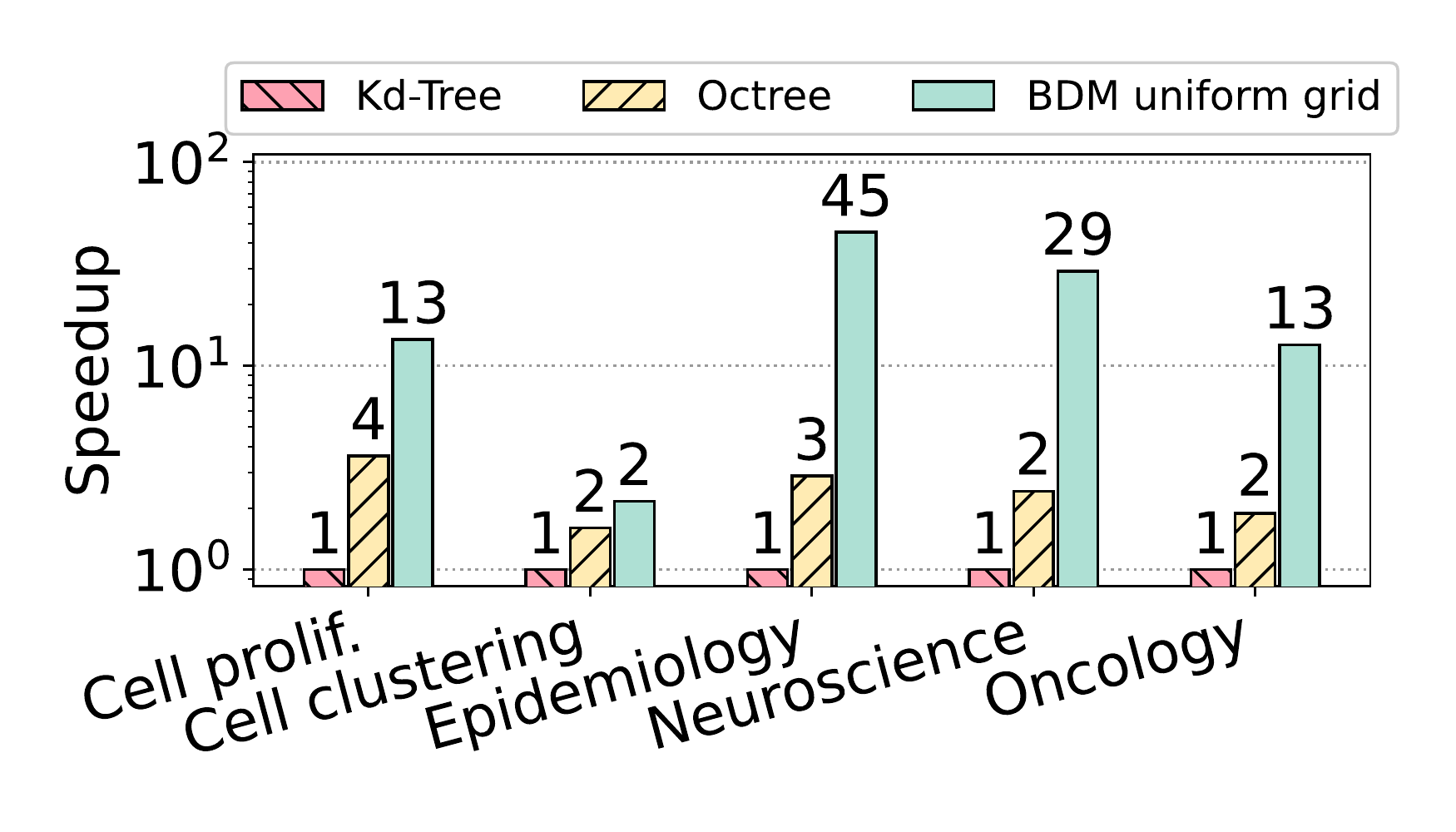}
     \caption{Whole simulation}
   \end{subfigure}
  \begin{subfigure}{\linewidth}
    \centering
    \includegraphics[width=0.49\textwidth]{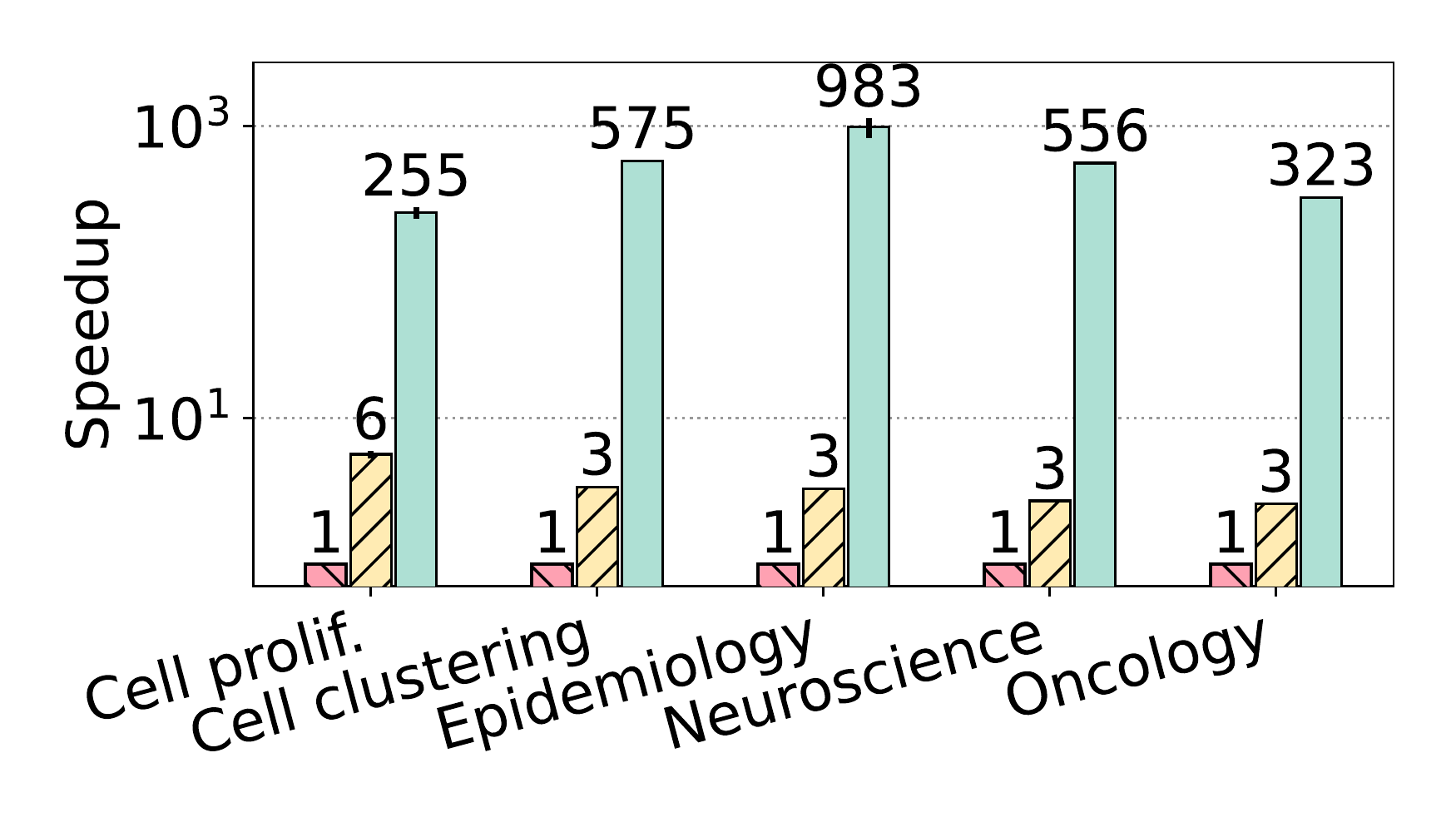}
    \includegraphics[width=0.49\textwidth]{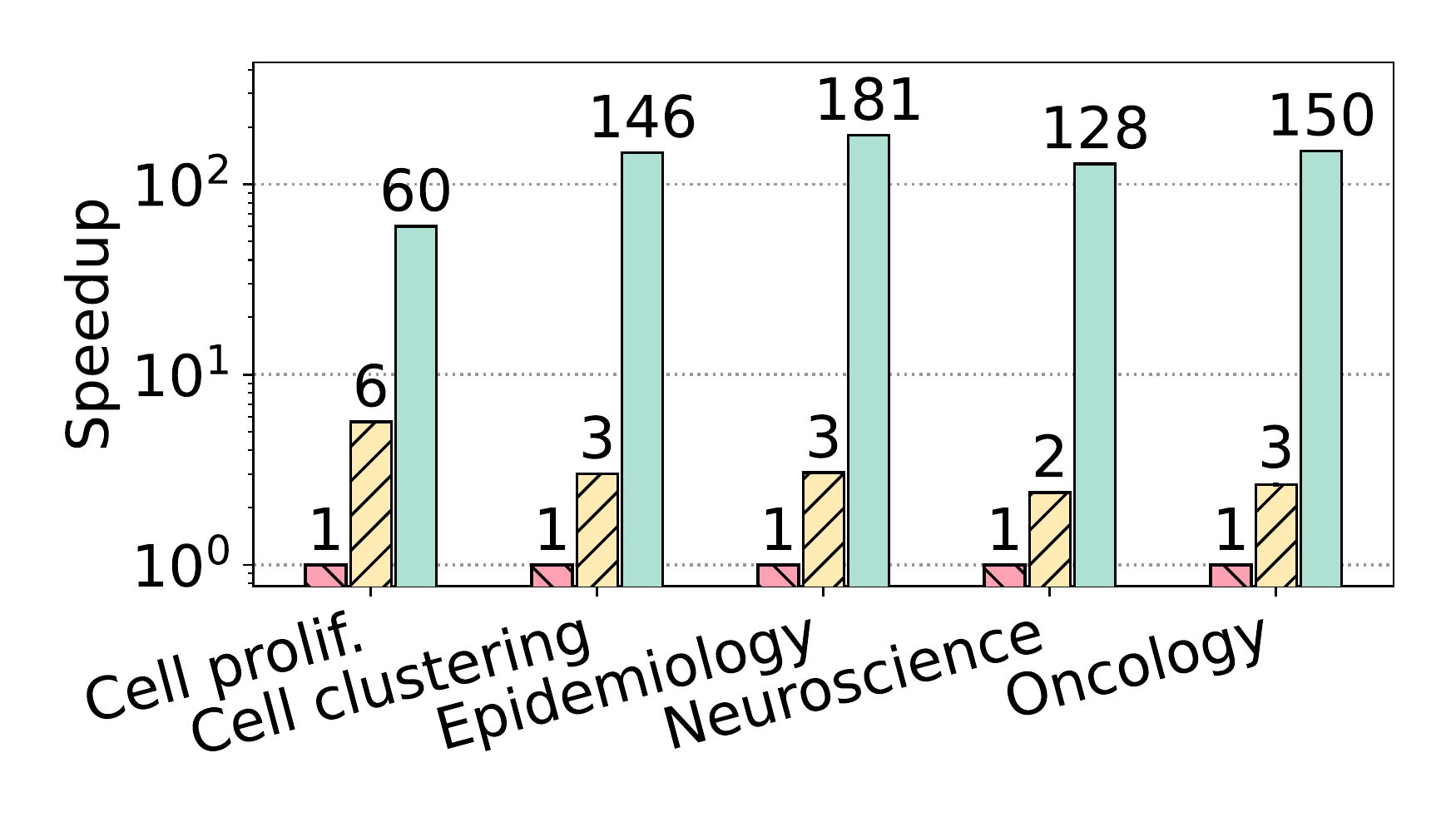}
     \caption{Build time}
   \end{subfigure}
  \begin{subfigure}{\linewidth}
    \centering
    \includegraphics[width=0.49\textwidth]{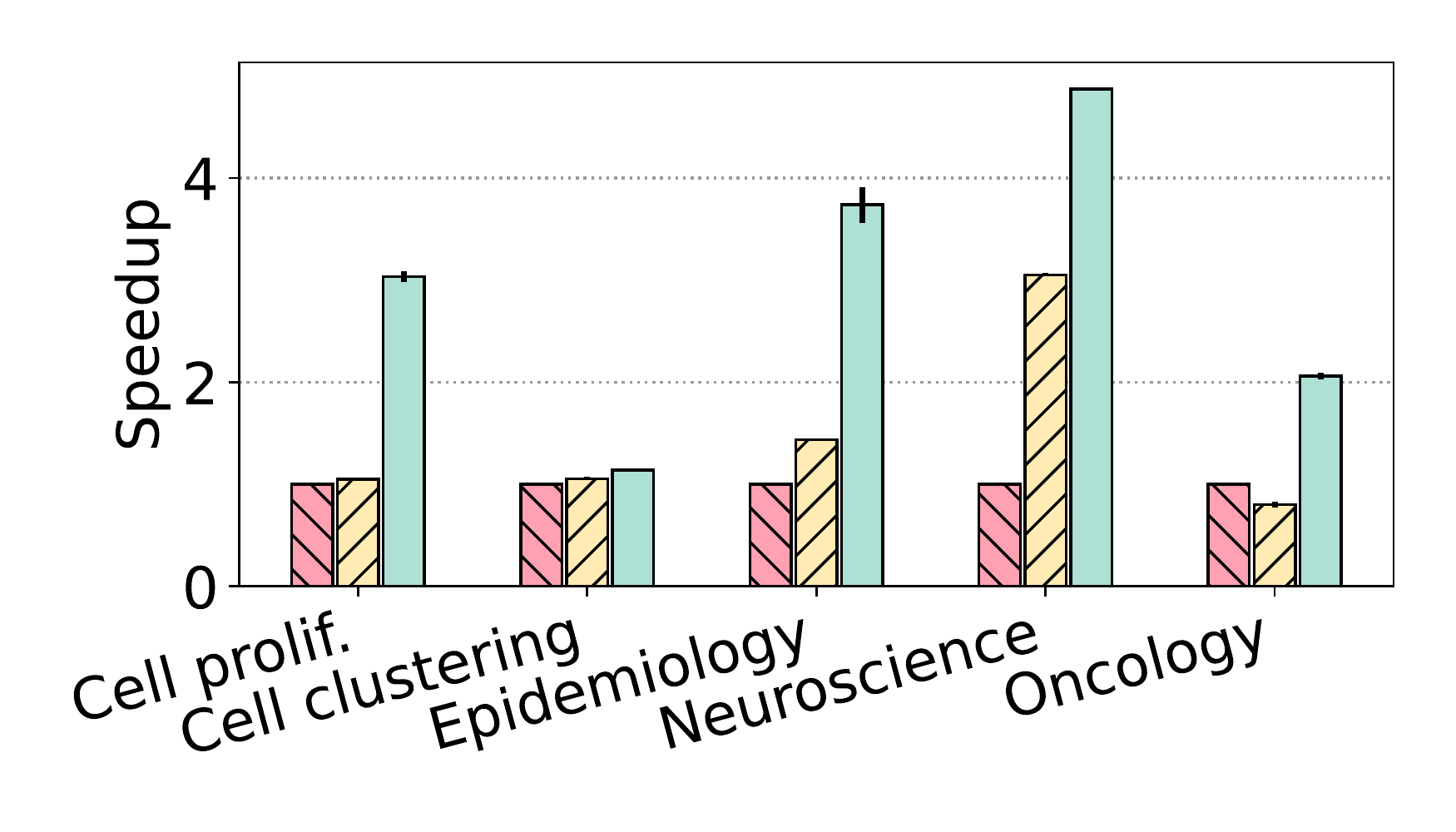}
    \includegraphics[width=0.49\textwidth]{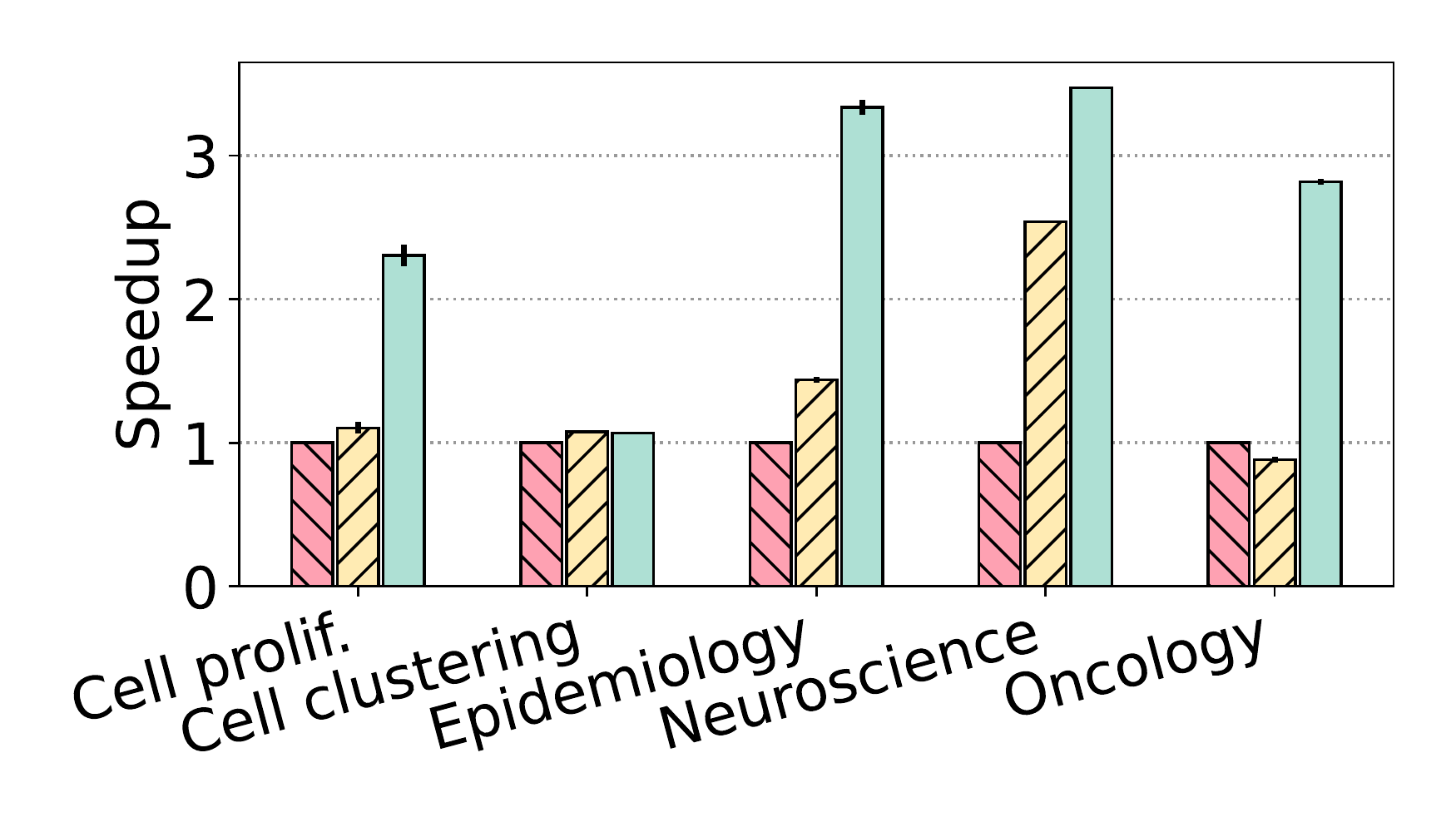}
     \caption{Search time (indirect)}
   \end{subfigure}
  \begin{subfigure}{\linewidth}
    \centering
    \includegraphics[width=0.49\textwidth]{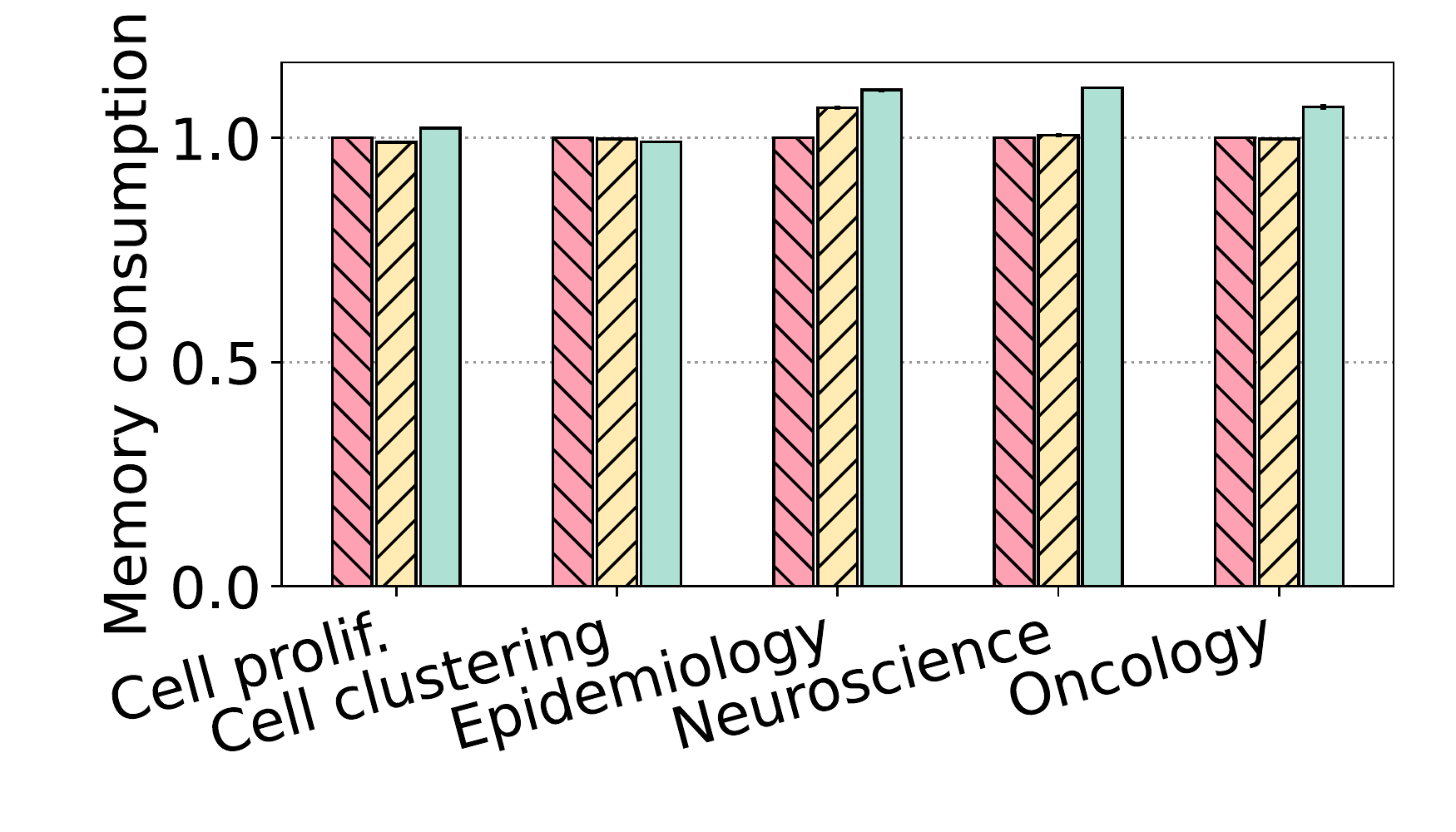}
    \includegraphics[width=0.49\textwidth]{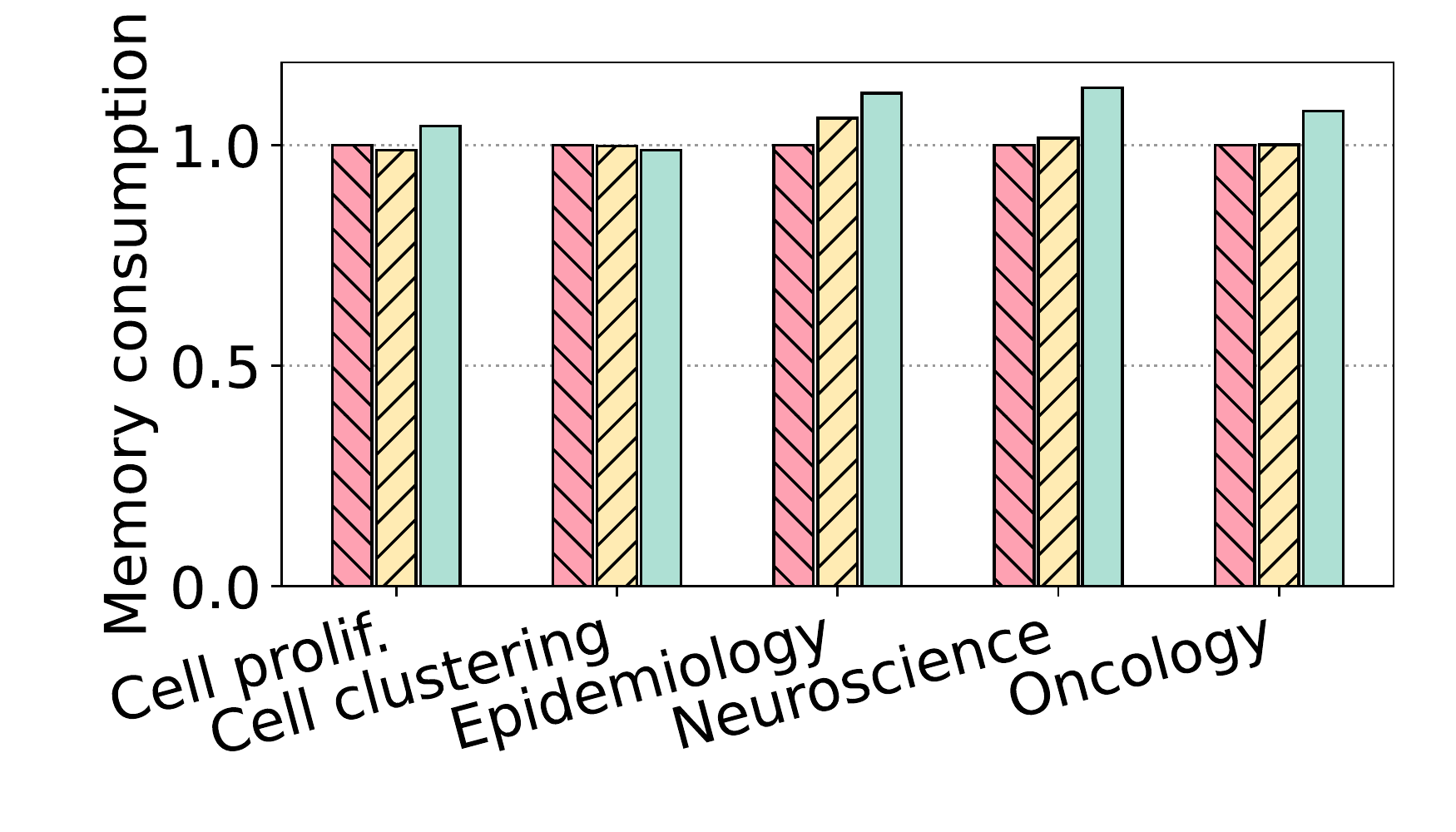}
    \caption{Memory consumption}
   \end{subfigure}
\caption{Neighbor search algorithm comparison
   (left column: four NUMA domains and 144 threads, right column: one NUMA
    domain
    and 18 threads). 
    The legend is shared between the plots.
  }
\label{fig:environment}
\end{figure}

Figure~\ref{fig:environment} compares three different neighbor search
  algorithms: \bdm{}'s uniform grid, UniBN's octree \cite{unibn}, and the kd-tree
  from nanoflann \cite{nanoflann}.
To ensure a fair comparison, we turned off agent sorting for all algorithms
  because it is currently only implemented for the uniform grid.
We validate our choice for the octree bucket size and nanoflann depth parameter
  and observe that the used parameters are within \result{4.20\%} of the optimum
  runtime.
The left column in Figure~\ref{fig:environment} shows the result for four NUMA
  domains and 144 threads, while the right column shows results for one NUMA
  domain and 18 threads.
We analyzed four properties of these radial neighbor search methods: runtime
  impact on the whole simulation, build and search time of the index, and memory
  consumption (Figure~\ref{fig:environment}).
We measure the search time indirectly by comparing the agent operation
  runtimes.
This operation contains the initial neighbor searches and thus provides
  information on how fast searches are executed.

The \bdm{} uniform grid implementation shows its benefits not only in the pure
  build time comparison but also in the full simulation analysis.
Although a significant build time difference in comparison to the kd-tree and
  octree is expected (because the build process is serial), the magnitude between
  \result{255 and 983$\times$} on four NUMA domains is surprising.
The uniform grid outperforms the other algorithms also during the search stage
  for all simulations.

Simulations using \bdm{}'s uniform grid implementation are up to
  \result{191$\times$} faster than the kd-tree implementation while consuming
  only \result{11\%} more memory in the worst case.

\subsection{NUMA-Aware Iteration}

We evaluate the individual performance impact of NUMA-aware iteration
(Section~\ref{sec:numa-iteration}).
In the other benchmarks, this optimization was included in the 
  ``memory layout optimization'' group.
We compare the simulation runtime with all optimizations enabled, 
  to executions in which ``NUMA-aware iteration'' is turned off.
This benchmark shows that this mechanism reduces the runtime between
  \result{1.07$\times$} and \result{1.38$\times$} (median:
  \result{1.30$\times$}).

\subsection{Agent Sorting and Balancing}

\begin{figure}[b]
  \centering
\begin{subfigure}{.49\linewidth}
    \includegraphics[width=\textwidth]{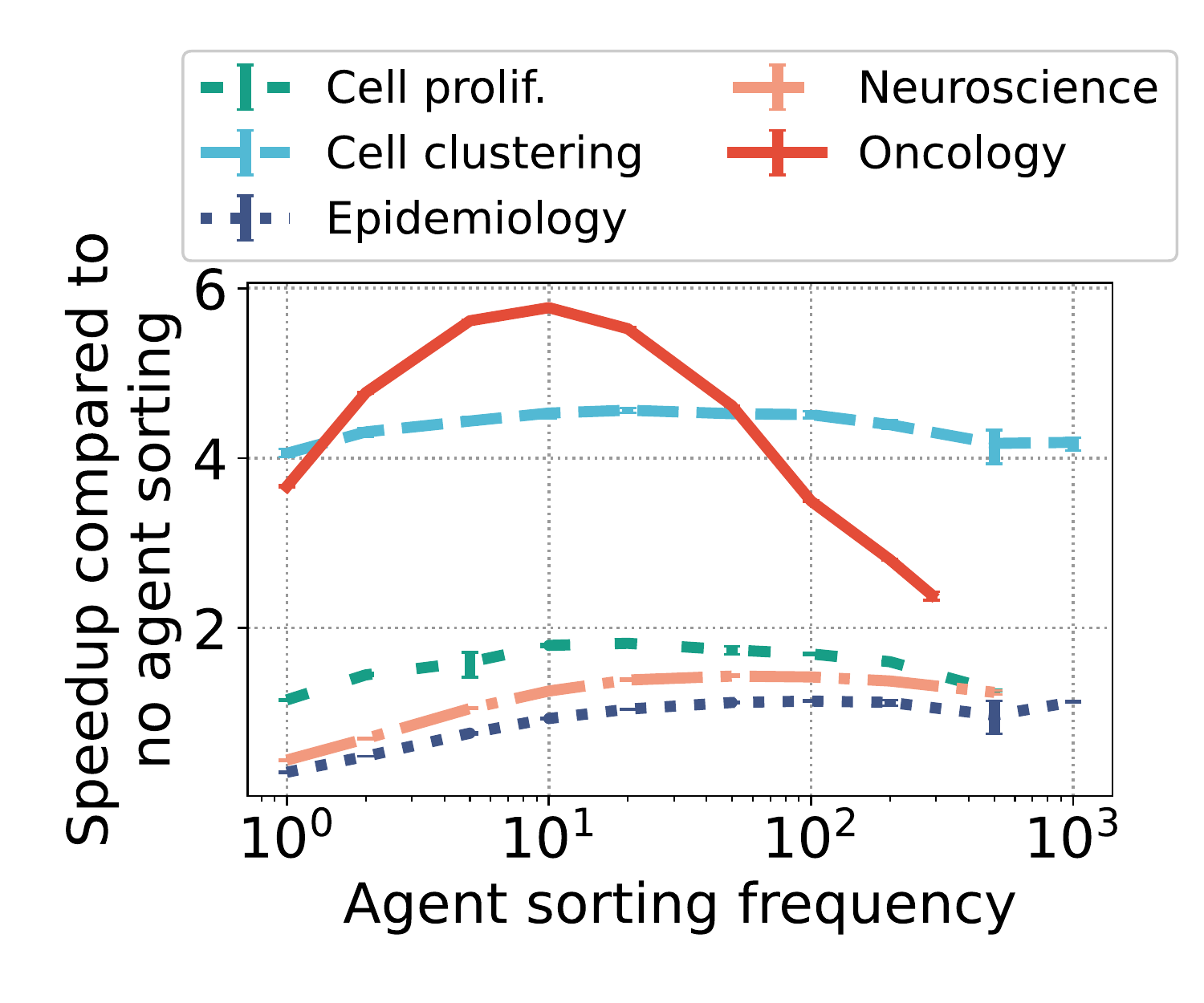}
\end{subfigure}
  \begin{subfigure}{.49\linewidth}
  \includegraphics[width=\textwidth]{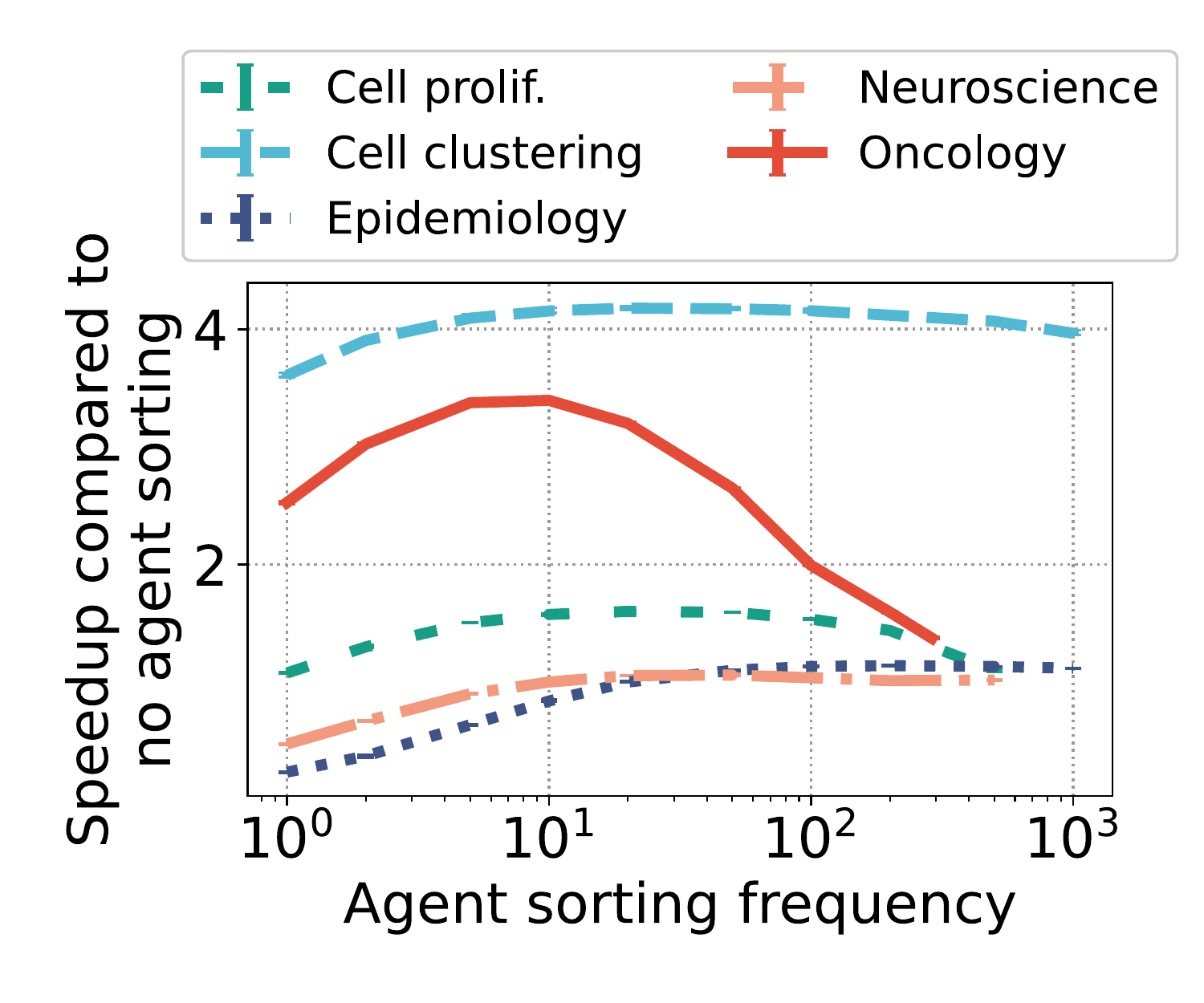}
\end{subfigure}
\caption{Agent sorting and balancing speedup for different execution
    frequencies (left: four NUMA domains and 144 threads, right: one NUMA
    domain
    and 18 threads).}
\label{fig:load-balancing}
\end{figure}

This section evaluates the impact of agent sorting and balancing
  (Section~\ref{sec:load-balancing}) on the simulation runtime for one and four
  NUMA domains.
To this extent, we perform a parameter study with varying agent sorting
  frequencies for each simulation.
Figure~\ref{fig:load-balancing} shows the speedup for four NUMA domains (left)
  and one NUMA domain (right).
The baselines in both cases are simulations without agent sorting.
An agent sorting frequency of one means that the operation is executed in every
  iteration; similarly, a frequency of ten would mean that the operation is
  executed every ten iterations.

Load balancing of agents among NUMA domains greatly impacts performance even on
  systems without NUMA architecture.
This stems from the fact that the agent sorting operation also aligns agents
  that are close in space also in memory.

The oncology and cell clustering simulations benefit most of this performance
  improvement (peak speedup of \result{5.77 and 4.56$\times$} for four NUMA
  domains).
Both simulations are initialized with a random distribution of agents.
Although the epidemiology simulation is also initialized randomly, its agents
  also move randomly with large distances between iterations.
This behavior reduces the alignment improvements significantly (peak speedup
  \result{1.14$\times$} for four NUMA domains).
The cell proliferation simulation is initialized with a 3D grid of cells, which
  improves the alignment compared to the worst-case random initialization.
Therefore, the maximum obtained speedup is reduced to \result{1.82$\times$}
  (four NUMA domains).
Suppose we change the initialization of the cell proliferation simulation to
  random, the maximum speedup increases to \result{4.68$\times$}.
This optimization performs below average for the neuroscience simulation.
This simulation only has an active growth front, while the remaining part
  remains static.
The static agent detection mechanism exploits this fact and avoids calculating
  mechanical forces for the static regions.
Therefore, the number of neighbor accesses is significantly reduced, and thus
  the benefits of aligned agents.
If static region detection is disabled, agent sorting and balancing improve the
  runtime by \result{3.80$\times$ at a frequency of 20}.

\subsection{\bdm{}
Memory Allocator} 

  To evaluate the performance of the \bdm{} memory allocator, we compare it with
  glibc's version of ptmalloc2 \cite{gloger2006ptmalloc} and jemalloc
  \cite{evans2011scalable} using our five benchmark simulations.
A comparison with tcmalloc \cite{tcmalloc} was impossible due to deadlock issues that we discovered during benchmarking.
Only the epidemiology use case uses additional memory during agent sorting and
  balancing.
Since the \bdm{} memory allocator only covers agents and behaviors, we need to
  use another allocator for the remaining objects.

\begin{figure}[h]
  \centering
  \begin{subfigure}{0.49\linewidth}
    \includegraphics[width=\textwidth]{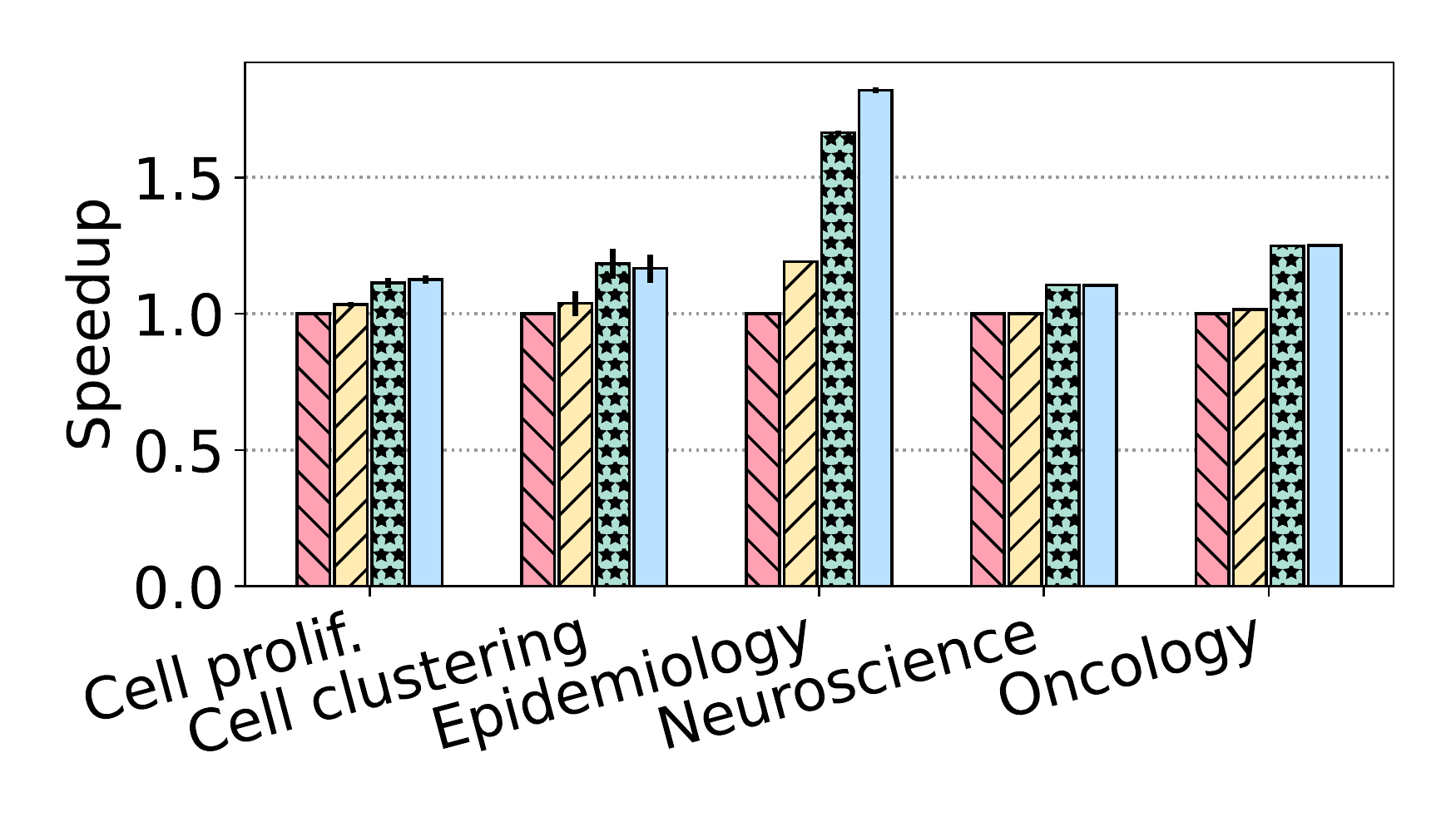}
\end{subfigure}
  \begin{subfigure}{0.49\linewidth}
    \includegraphics[width=\textwidth]{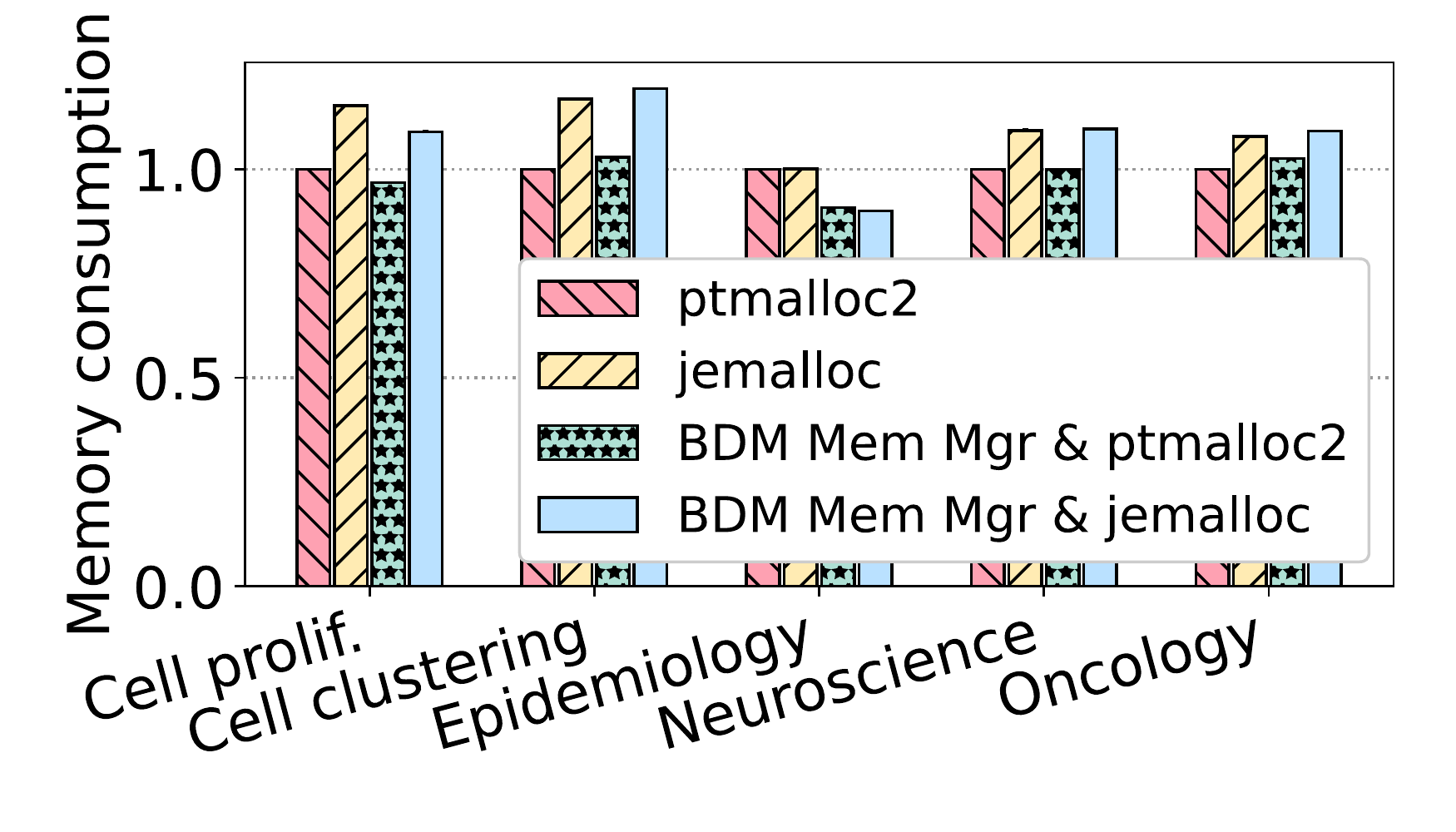}
\end{subfigure}
\caption{Memory allocator comparison (left: speedup, right: memory consumption). The legend is shared between the plots.}
\label{fig:mem-mgr-pc}
\end{figure}

This requirement results in four tested configurations per simulation, as
  illustrated in Figure~\ref{fig:mem-mgr-pc}.
The \bdm{} memory allocator improves the overall simulation runtime up to
  \result{1.52$\times$} over ptmalloc2 (\result{median: 1.19$\times$}) and up to \result{1.40$\times$} over jemalloc
  (\result{median 1.15$\times$}).
The allocator consumes \result{1.41\%} less memory than ptmalloc2 and \result{2.43\%} less memory than
  jemalloc on average.

\section{Related Work}
\label{related-work}

\paragraph{Agent-based simulation tools}
To our knowledge, this is the first paper to present an agent-based simulation
  engine capable of simulating neuroscientific models with billions of agents.
Biocellion \cite{biocellion} and Timothy \cite{cytowski_large-scale_2014} can
  also simulate one billion agents, but neither of them support simulations 
  in the demanding computational neuroscience domain.
Our performance evaluation shows that \bdm{} is \result{9.64$\times$} more 
  efficient than Biocellion \cite{biocellion} for a simulation with 1.72 billion agents.
Furthermore, \bdm{} is three orders of magnitude
  faster than the popular neuroscientific simulator Cortex3D
  \cite{zublerdouglas2009framework}, and the general-purpose tool NetLogo \cite{netlogo} on our benchmark hardware.
Other tools \cite{chaste, physicell, richmond_high_2010} also support
  large-scale models, but only show simulations of up to $10^6$  agents.

\paragraph{Memory layout optimizations}
Data movement between main memory and processor cores is a fundamental
  bottleneck in today's computing systems \cite{mutlu_processing_2019}.
Recent research in computer architecture explores new approaches to address
  this bottleneck, such as processing-in-memory, i.e., placing compute capability
  closer to the data~\cite{gomez-luna_benchmarking_2021, ahn_pim-enabled_2015}.
Agent-based simulation tools are also negatively impacted by the data movement
  bottleneck.
We address this problem in software with a better memory layout resulting in
  more efficient bandwidth utilization and data reuse in caches.
Space-filling curves \cite{morton1966computer, hilbert_ueber_1891,
    peano_sur_1890} can improve the memory layout by aligning objects that are
  close in 3D space.
Therefore, these curves are frequently used to optimize geometric data
  structures \cite{unibn, asano_space-filling_1997} and molecular dynamics
  simulations \cite{grime_highly_2014, anderson_general_2008,
    nakano_scalable_1999}.
To our knowledge, none of the other agent-based simulation frameworks (e.g., \cite{physicell, biocellion, mason, chaste, netlogo, zublerdouglas2009framework, agentcell, bsim2, netmorph, idynomics, torben-nielsen_context-aware_2014, morpheus, cytowski_large-scale_2015, dmason, repast-simphony, repast-hpc, richmond_high_2010, CompuCell3D}) use space-filling curves to improve the cache hit rate and minimize the amount of remote DRAM accesses. 
We introduce this proven technique to the agent-based workload and present a
  mechanism to determine the Morton order of a non-cubic grid in linear time.

\paragraph{Neighbor search}
Agent-based simulation platforms use various neighbor search algorithms:
  Delaunay triangulation \cite{zublerdouglas2009framework}, octree
  \cite{hauri_self-construction_2013}, and grid-based approaches \cite{netlogo,
    richmond_high_2010, biocellion}.
Grids are also commonly used on the GPU \cite{hoetzlein_fast_2014, unibn,
    gross_fast_2019, aaby_efficient_2010, dsouza}.
Depending on the dataset and specific search query ([fixed-]radius neighbor
  search or k-nearest neighbors), the literature recommends different algorithms
  \cite{vermeulen_comparative_2017, unibn}.
Our contribution lies in the efficient implementation and integration of the
  uniform grid into the simulation engine and in providing insights into the
  performance differences for the agent-based workload.

\paragraph{Performance evaluation}
To our knowledge, this paper presents the most comprehensive performance
  analysis of an agent-based simulation platform.
Existing platforms report only limited performance results, including
  simulation execution times and occasionally scalability analyses
  \cite{zublerdouglas2009framework, chaste, biocellion, repasthpc, mason,
    physicell}.
Performance data can also be found in model papers
  \cite{strazdins_performance_2011, murphy_simulating_2016} and in  works that
  focus on hardware accelerators \cite{xiao_survey_2019}.
We improve upon these works by providing an in-depth analysis of each
  performance-relevant component.
Efforts in the direction of a standard agent-based benchmark have been made by
  Moreno et al.
\cite{moreno_designing_2019} and
Rousset et al. \cite{rousset_survey_2016}.
However, these synthetic benchmarks fall short of representing a realistic
  range of agent-based simulations by over-simplifying memory access patterns and
  assuming that agents always move randomly.
Compared to these, our benchmark simulations cover a broader spectrum of
  performance relevant simulation metrics (see
  Table~\ref{tab:sim-characteristics}).

\paragraph{Comparison outside the ABM field}
Other particle-based applications, such as molecular dynamics (MD) \cite{phillips_scalable_2005, thompson_lammps_2022, vasp}, astrophysics (AP) \cite{spiridon_n-body_2013}, or computational fluid dynamic (CFD) \cite{jasak_openfoam_2009} simulations often face similar computational challenges to improve the performance of large-scale simulations. 
LAMMPS \cite{thompson_lammps_2022}, for example, also uses a grid-based structure to determine neighbors. 
While LAMMPS stores neighbor lists for each atom, which according to Thompson et al. \cite{thompson_lammps_2022} ``consumes the most memory of any data structure in LAMMPS'', \bdm{} does not need these lists and therefore saves memory.
\bdm{} improves over LAMMPS and VASP \cite{vasp} by sorting agents using a space-filling curve 
  (Section~\ref{sec:load-balancing}) and using a custom memory allocator (Section~\ref{sec:mem-allocator}) to reduce the memory access latency.
A NUMA-aware thread allocation mechanism,
as the one used in \bdm{} (Section~\ref{sec:numa-iteration}),
is not needed in LAMMPS or VASP because both tools support distributed parallelism with MPI.
In this work, we identify several computational challenges in ABM, 
which we tackle by using methods inspired by MD, AP, and CFD.
The main difference between ABM and other particle-based applications is that the computations can vary significantly from each other in terms of arithmetic intensity, the number of considered neighbors, data access patterns, and more, thus posing diverse computational challenges.

\section{Conclusion and Future Work}

This paper presents a novel agent-based simulation engine optimized for high
  performance and scalability.
\bdm{} enables not only larger-scale simulations, but also helps researchers of
  small scale studies with accelerated parameter space exploration, and faster
  iterative development.

We identify general agent-based performance challenges and provide six
  solutions to maximize parallelization, reduce memory access latency and data
  transfers, and avoid unnecessary work.
These solutions are transferable and can be used to accelerate other
  agent-based simulation tools.

We present a comprehensive performance analysis to provide insights into the
  agent-based workload and to give our users a better understanding of \bdm{}'s
  capabilities.
We find that on our system, the presented optimizations improve performance up
  to \result{524$\times$ (median 159$\times$)} and allow \bdm{} to scale to 72
  physical processor cores with a parallel efficiency of \result{91.7\%}.
A comparison with state-of-the-art tools shows that \bdm{} is up to three
  orders of magnitude faster.
These performance characteristics enable simulations with billions of agents,
  as demonstrated in our analysis.

Our performance optimizations, which are effective on machines with one or more NUMA domains,  are an important stepping stone towards a distributed simulation engine with a 
hybrid MPI/OpenMP design. 
Ongoing work focuses on realizing this distributed simulation engine capable of dividing
  the computation among multiple nodes to push the boundaries of agent-based
  simulation even further.
 
\begin{acks}
  We want to thank the CERN Knowledge Transfer office (\url{https://kt.cern/}) for supporting this work.
  We acknowledge the support provided to the SAFARI Research Group by our industrial partners, including Huawei, Intel, Microsoft, and VMware.
\end{acks}

\bibliographystyle{ACM-Reference-Format}
\interlinepenalty=10000

\appendix

\section{Artifact Description}
This appendix contains a short summary of the instructions to reproduce the results in the paper. 
The whole process is fully automated and generates all plots and visualizations shown.
The complete instructions can be found in the \texttt{SF1-readme.pdf} file located at \url{https://doi.org/10.5281/zenodo.6463816} \hspace{1mm} and 
\hspace{1mm} \url{https://github.com/CMU-SAFARI/BioDynaMo}.

\vspace{2ex}
\noindent
\textbf{List of Files:} We provide the following supplementary files on Zenodo (\url{https://doi.org/10.5281/zenodo.6463816}) as well as the SAFARI Research Group's Github page \url{https://github.com/CMU-SAFARI/BioDynaMo}:
\begin{itemize}
\item \texttt{SF1-readme.pdf}: This document provides detailed documentation, step-by-step instructions to set up the systems and execute the benchmark scripts, and ideas on how to reuse and repurpose this artifact.
\item \texttt{SF2-code.tar.gz}: This archive contains all the necessary code to produce all results shown in the paper.
\item \texttt{SF3-bdm-publication-image.tar.gz}: This archive contains a self-contained docker image to simplify executing our benchmarks and aid long-term reproducibility.
\item \texttt{SF4-raw-results.tar.gz}: This archive contains the raw results we obtained when we executed the benchmarks on our systems.
\end{itemize}

\subsection{Getting Started}
Setting up a new system requires four steps.
More details can be found in Section~2 in \texttt{SF1-readme.pdf}.

\begin{enumerate}
  \item Download and extract the code archive in Supplementary File SF2.
  \item Install the following software packages on the host machine: 
    Docker (version $>= 19.0.3$), 
    Intel Vtune sampling driver (version 2022.2.0),
    and the Linux \texttt{screen} command.
  \item Load the docker image provided by Supplementary File SF3.
  \item Verify the setup by executing the following command in the \texttt{bdm-paper-examples} directory:
    \texttt{docker/run.sh ./run-functional-evaluation.sh}.
\end{enumerate}

\subsection{Reproducing Results}

We separate the benchmarks into multiple scripts with different memory and disk space requirements to allow researchers with less powerful hardware to execute a subset of benchmarks.
To execute one of the scripts below, change into the \texttt{bdm-paper-examples} directory, and pass the script as parameter to the command \texttt{docker/run.sh}.
More details can be found in Section~4 in \texttt{SF1-readme.pdf}.

\begin{itemize}

  \item \texttt{run-main.sh}:
    This script executes the majority of benchmarks described in the evaluation section of the
    paper and outputs Figure~\ref{fig:operation-breakdown-uarch-analysis} (left) and Figures~\ref{fig:optimization-overview}--\ref{fig:mem-mgr-pc}.

\item \texttt{run-comparison-with-others.sh}:
This script executes the comparison of BioDynaMo with Cortex3D and NetLogo and outputs 
Figure~\ref{fig:comparison-with-others}.

\item \texttt{run-runtime-complexity.sh}:
This script analyses the runtime and memory consumption of BioDynaMo, with the number
of agents increasing from $10^3$ to $10^9$, and generates
Figure~\ref{fig:runtime-complexity}.

\item \texttt{run-profiling.sh}:
This script performs the microarchitecture analysis for all simulations in Table~\ref{tab:sim-characteristics} 
  and generates
Figure~\ref{fig:operation-breakdown-uarch-analysis} (right).

\item \texttt{run-biocellion-cmprsn-single-node.sh}: 
This script executes the small-scale comparison with Biocellion and the optimization
    analysis in Figure~\ref{fig:biocellion-comparison} (right).

\item \texttt{run-biocellion-cmprsn-cluster.sh}: 
This script executes the large-scale comparison with Biocellion, the optimization analysis in
    Figure~\ref{fig:biocellion-comparison} (left), and renders the visualization in Figure~\ref{fig:biocellion-rendering}.
\end{itemize}

\subsection{Reusing and Repurposing the Artifact}
Besides reproducing the results, researchers can build upon our artifact in numerous ways.
A non-exhaustive list of possibilities with detailed instructions is given in Section~5 in \texttt{SF1-readme.pdf}.
These possibilities include:

\begin{itemize}
  \item Add additional benchmarks
  \item Evaluate the effectiveness of additional optimizations
  \item Evaluate BioDynaMo’s performance for additional simulations
\end{itemize}

\subsection{Contact}
Please contact us with any feedback or if you need any help building on \bdm{}.
You can reach us at \href{mailto:lukas.breitwieser@gmail.com}{lukas.} \href{mailto:lukas.breitwieser@gmail.com}{breitwieser@gmail.com} and \href{mailto:omutlu@gmail.com}{omutlu@gmail.com}.
 
\end{document}